\documentclass[12pt, a4paper]{article}
\usepackage{cite}
\usepackage{amsmath,amssymb}
\input{colordvi.tex}
\usepackage{pifont}
\usepackage{subfigure}
\usepackage{comment}
\usepackage{bm}
\usepackage{url}
\usepackage{braket}

\usepackage[linktocpage]{hyperref}
\usepackage{ytableau}
\everymath{\displaystyle}
\usepackage{cancel}
\usepackage{mathtools}
\usepackage{tabularx}
\usepackage{lipsum}
\usepackage[bottom]{footmisc}

\long\def\symbolfootnote[#1]#2{\begingroup%
  \def\thefootnote{\fnsymbol{footnote}}\footnote[#1]{#2}\endgroup}

\usepackage{ifpdf}
\ifpdf
  \usepackage{graphicx, hyperref, xcolor}     
\else     
  \usepackage[dvipdfmx]{graphicx, hyperref, xcolor}     
 \fi

\newcommand{\be}{\begin{equation}}
\newcommand{\ee}{\end{equation}}
\newcommand{\ba}{\begin{eqnarray}}
\newcommand{\ea}{\end{eqnarray}}

\definecolor{blue}{rgb}{0, 0.4, 0.7}
\definecolor{Blue}{rgb}{0, 0, 1}
\hypersetup{
setpagesize=false,
bookmarksnumbered=true,%
bookmarksopen=true,%
colorlinks=true,%
linkcolor=blue,
urlcolor=blue,
citecolor=blue,
}

\DeclareUnicodeCharacter{2212}{-}
\begin{document}
\fontsize{12pt}{18pt}\selectfont

\begin{titlepage}

\begin{center}

\hfill CERN-TH-2022-149\\

\vskip .5in

{\Large \bf
High Quality Axion in Supersymmetric Models}

\vskip .5in

{\large
Gongjun Choi$^{(a)}$\symbolfootnote[1]{gongjun.choi@cern.ch} and Tsutomu T. Yanagida$^{(b,c)}$\symbolfootnote[2]{tsutomu.tyanagida@sjtu.edu.cn}
}
\vskip 0.25in

$^{(a)}${\em Theoretical Physics Department,
      CERN},\\
      {\em 1211 Geneva 23, Switzerland}\\[.3em]
\vskip 0.1in

$^{(b)}${\em Tsung-Dao Lee Institute (TDLI)},\\
{\em \& School of Physics and Astronomy, Shanghai Jiao Tong University},\\
      {\em Shengrong Road 520, 201210 Shanghai, P.\ R.\ China}\\[.3em]
\vskip 0.1in

$^{(c)}${\em Kavli IPMU (WPI), The University of Tokyo},\\
{\em Kashiwa, Chiba 277-8583, Japan}\\[.3em]
\vskip 0.20in

\end{center}
\vskip .2in

\begin{abstract}
In this work, we discuss how the use of the symmetries well motivated in physics beyond the Standard model (BSM) can guarantee the high quality axions. We avoid to introduce symmetries only useful for addressing the axion quality problem. Rather, we rely on symmetries well motivated by other issues in BSM: supersymmetry, $U(1)_{\rm B-L}$ and the discrete R-symmetry $Z_{NR}$. We show that the interplay among these guarantees the high quality of the axion even for the gravitino mass and axion decay constant as large as $m_{3/2}=\mathcal{O}(10){\rm TeV}$ and $F_{a}=\mathcal{O}(10^{15}){\rm GeV}$ respectively. The key point of this work relies on the observation that the MSSM contribution to the mixed anomalies $Z_{NR}-[SU(2)_{L}]^{2}$ and $Z_{NR}-[SU(3)_{c}]^{2}$ is not enough for gauging $Z_{NR}$ for $N\neq6$, which necessitates the introduction of new matter fields. We make the introduction to achieve zero mixed anomalies, which logically supports a desired large enough $N$ for $Z_{NR}$. This mechanism effectively makes $Z_{NR}$ equal to $U(1)_{R}$ and thus offers a logically complete solution to the axion quality problem.

\end{abstract}

\end{titlepage}


\renewcommand{\thepage}{\arabic{page}}
\setcounter{page}{1}
\renewcommand{\thefootnote}{\arabic{footnote}}
\setcounter{footnote}{0}
\renewcommand{\theequation}{\thesection.\arabic{equation}}


\newpage

\tableofcontents

\newpage

\section{Introduction}
\label{sec:Intro}
\setcounter{equation}{0}

QCD axion is one of the most well-motivated hypothetical particles in physics beyond the Standard model (BSM) as it dynamically resolves the long standing strong CP problem~\cite{Peccei:1977hh,Peccei:1977ur,Weinberg:1977ma,Wilczek:1977pj}. It is a pseudo Nambu-Goldstone boson (pNGB) arising from the spontaneous breaking of the global $U(1)_{\rm PQ}$ symmetry which has the mixed anomaly with $SU(3)_{c}$. As such, it couples to the QCD anomaly term via the operator 
\be
\mathcal{L}\supset\frac{a}{F_{a}}\frac{g_{s}^{2}}{32\pi^{2}}G_{\mu\nu}^{b}\tilde{G}^{b\mu\nu}\,,
\label{eq:anomalyterm}
\ee
where $F_{a}$ is an axion decay constant, $g_{s}$ is the gauge coupling for $SU(3)_{c}$, $G_{\mu\nu}^{a}$ is the $SU(3)_{c}$ gauge field strength ($b$ is the group generator index), and $\tilde{G}^{a\mu\nu}$ is the dual to $G_{\mu\nu}^{a}$. Introduction of the coupling in Eq.~(\ref{eq:anomalyterm}) renders the $\theta$-parameter for QCD vacua a dynamical variable and the potential for $\bar{\theta}\equiv\theta+(a/F_{a})$ generated by the non-perturbative effects of the QCD vacuum has the minimum at $\bar{\theta}=0$~\cite{Vafa:1983tf}. Therefore, we can understand the experimental constraint on $\bar{\theta}\lesssim10^{-10}$~\cite{Baker:2006ts} from the measurement of the neutron electric dipole moment with the aid of the dynamical relaxation of the axion field toward $\langle a/F_{a}\rangle+\theta=0$.

This elegant Peccei-Quinn mechanism to solve the strong CP problem, however, becomes challenged by potential modifications $\Delta V(\theta+\delta)$ to $V(\theta)$ with $\delta$ a phase shift. This change in general causes the shift in the global minimum of the axion potential $\Delta\bar{\theta}_{\rm min}$, which spoils the Peccei-Quinn mechanism.

One of the potential sources for $\Delta V(\theta+\delta)\neq0$ is a $U(1)_{\rm PQ}$ violating higher dimensional operator like $c^{(n)}_{\Phi}(\Phi^{n}+\Phi^{\dagger\, n})/M_{P}^{n-4}$ ($n\geq5$) where $\Phi=(\phi/\sqrt{2})e^{ia/F_{a}}$ is the PQ scalar, $c_{\Phi}^{(n)}$ a complex dimensionless coupling constant and $M_{P}\simeq2.4\times10^{18}{\rm GeV}$ is the reduced Planck mass~\cite{Kamionkowski:1992mf,Holman:1992us,Barr:1992qq,Ghigna:1992iv}. Given $\Delta\bar{\theta}_{\rm min}\simeq\Delta V(\theta+\delta)/(m_{a}^{2}F_{a}^{2})$ with $m_{a}F_{a}\simeq\Lambda_{\rm QCD}^{2}\simeq(0.2{\rm GeV})^{2}$, it is realized that axion can still be a good solution to the strong CP problem only if $\Delta V(\theta+\delta)$ can be sufficiently suppressed to give $\Delta\bar{\theta}_{\rm min}<10^{-10}$. This problem of suppressing $\Delta V(\theta+\delta)$ is referred to as {\it axion quality problem}. 

Unless the dangerous higher dimensional operators are suppressed with extremely small coefficients $c_{\Phi}^{(n)}<\!\!<1$, the aforesaid operator contributes to $\Delta\bar{\theta}_{\rm min}$ by
\be
\Delta\bar{\theta}_{\rm min}\simeq10^{-10}\times10^{86}\times\left(\frac{F_{a}}{M_{P}}\right)^{n},\qquad(n\geq5)\,.
\label{eq:thetashift}
\ee
Now that there is no any definite theoretical prediction for $F_{a}$, in principle any $F_{a}$ value greater than $10^{9}{\rm GeV}$ coming from the stellar cooling process~\cite{Raffelt:2006cw} is allowed. And thus Eq.~(\ref{eq:thetashift}) tells us that a larger $F_{a}$ causes a much larger $\Delta\bar{\theta}_{\rm min}$ when there is no further suppression in coefficients of operators.\footnote{As a matter of fact, some of higher dimensional operators are generated via non-pertuabative gravitational effects and in that case, the coefficients  are given by $\mathcal{O}(e^{-S_{\rm grav}})$ where $S_{\rm grav}\sim M_{P}/F_{a}$ is a gravitational instanton action~\cite{Lee:1988ge,Giddings:1987cg,Kallosh:1995hi,Alonso:2017avz} (for the recent review on this, see \cite{Alvey:2020nyh}). Even in this case, a larger $F_{a}$ gives a larger coefficient and thus generally the larger $F_{a}$ makes the axion quality worse. For $F_{a}\geq2\times10^{16}{\rm GeV}$, $\Delta\bar{\theta}_{\rm min}>>10^{-10}$ becomes the case.} 

Often there arise axions with $F_{a}=\mathcal{O}(10^{16}){\rm GeV}$ from string theories as the Kaluza-Klein (KK) zero mode of higher form gauge fields~\cite{Svrcek:2006yi,Banks:2003sx,Conlon:2006tq,Acharya:2010zx}. If there exists a mechanism to avoid to have higher dimensional operators in the theory, it seems that we can keep the theoretically well-motivated QCD string axions as the solution to the strong CP problem. Then, how could we guarantee the absence of dangerous higher dimensional operators? Going a step further, would it be possible to achieve it with symmetries well-motivated by problems in BSM physics other than the strong CP problem?\footnote{Of course, one can impose an additional discrete $Z_{N}$ gauge symmetry with a very large $N\geq\mathcal{O}(10)$ under which the PQ scalar is charged. However, as $Z_{N}$ should be gauged, one needs to care about the mixed anomaly free conditions for $Z_{N}$, which makes introducing the gauged $Z_{N}$ non-trivial as we will see.}

In this work, motivated by these questions, we give our special attention to the supersymmetric extension of the Standard model (SSM) armed with the gauged $U(1)_{\rm B-L}$ symmetry and gauged discrete R-symmetry $Z_{NR}$. The anomaly free $U(1)_{\rm B-L}$ gauge symmetry is very well motivated in the context of the seesaw mechanism for explaining the tiny active neutrino masses~\cite{Minkowski:1977sc,Yanagida:1979as,PhysRevD.20.2986,GellMann:1980vs} and the leptogenesis~\cite{Fukugita:1986hr}. In addition, any model embedded in supergravity (SUGRA) enjoys the fundamental gauged $R$-symmetry.\footnote{Recently the use of discrete $R$-symmetry for addressing the axion quality problem was discussed in \cite{Baer:2018avn,Bhattiprolu:2021rrj}.} In this set-up, $U(1)_{\rm PQ}$ is understood to be the accidental remnant of $U(1)_{\rm B-L}$ gauge symmetry and most of the higher dimensional operators are suppressed simply because of gauge invariance of $U(1)_{\rm B-L}$. There remain only few dangerous $\mathcal{O}^{d\geq5}_{\cancel{\rm PQ}}$'s which can be further suppressed due to the discrete $Z_{NR}$ symmetry.\footnote{A gauge symmetry-assisted high quality axion was also discussed in \cite{Cheng:2001ys,Hill:2002kq,Babu:2002ic,Dias:2002gg,Harigaya:2013vja,Fukuda:2017ylt,Fukuda:2018oco,DiLuzio:2017tjx,Duerr:2017amf,Bonnefoy:2018ibr,Choi:2020vgb,Ardu:2020qmo,DiLuzio:2020qio,Kawamura:2020jzb,Darme:2021cxx,Chen:2021haa,Bhattiprolu:2021rrj,Berezhiani:1985in,Berezhiani:1990wn,Berezhiani:1989fp,Berezhiani:1992rk}. Other solutions include, for example, composite axion scenarios~\cite{Kim:1984pt,Choi:1985cb,Randall:1992ut,Redi:2016esr,Lillard:2017cwx,Lillard:2018fdt,Lee:2018yak,Gavela:2018paw,Vecchi:2021shj,Contino:2021ayn} and heavy axion scenarios~\cite{Dimopoulos:1979pp,Tye:1981zy,Rubakov:1997vp,Berezhiani:2000gh,Hook:2014cda,Fukuda:2015ana,Gherghetta:2016fhp,Valenti:2022tsc}.} 

The outline of the paper is what follows. In Sec.~\ref{sec:twosym}, taking a conservative attitude, we discuss how the two gauge symmetries motivated by other problems in BSM than the strong CP problem can help us achieve a high quality of axion. Sec.~\ref{sec:app} is dedicated to the presentation of an exemplary model and the model's prediction for the axion quality. We also show that the model does not suffer from the small size instanton-induced modification to the axion potential. In Sec.~\ref{sec:tension}, we discuss the upper bound of $F_{a}$ for which our solution to the axion quality problem is valid in light of the measurement for the abundance of primordial light elements. 

From here on, the same notation for a chiral superfield and its scalar component will be used. We will denote the R-charge of an operator $\mathcal{O}$ by $R[\mathcal{O}]$, and the charges of $U(1)_{\rm B-L}$ and $U(1)_{\rm PQ}$ by $Q_{\rm B-L}[\mathcal{O}]$ and $Q_{\rm PQ}[\mathcal{O}]$ respectively.

\section{Useful Symmetries}
\label{sec:twosym}
\setcounter{equation}{0}
In this section, we discuss additional symmetries we assume on top of the SM gauge group and supersymmetry, and how the assumption helps us achieve the high quality QCD axion. As we emphasized in the introduction, the aim of this work is to investigate if the symmetries suggested in BSM physics to address well-known problems other than the strong CP problem can be useful in addressing the axion quality problem. 

With such a purpose, we attend to three symmetries in this work: local supersymmetry, $U(1)_{\rm B-L}$ gauge symmetry and $R$-symmetry. Above all, the first one is one of the most attractive resolutions to the hierarchy problem as it allows for cancellation among radiative corrections to the Higgs mass. In addition to this, there are a number of advantages that supersymmetric theories enjoy including the radiatively-induced electroweak symmetry breaking, the possibility of the grand unification of the SM gauge couplings and so on.

In the following subsections, we study the rest of two additional symmetries in relation to the axion quality problem. The introduction of the matter contents and charge assignments based on the anomaly free conditions for the two gauge symmetries will offer us the logical reasoning for suppressing unwanted higher dimensional operators.

\subsection{$U(1)_{\rm B-L}$ Gauge Symmetry}
\label{sec:twoU(1)s}
Extending the particle contents of MSSM by the three right-handed (RH) neutrinos carrying the opposite lepton number to that of the active neutrinos can provide us with the explanation for the tiny masses of the active neutrinos via the seesaw mechanism~\cite{Minkowski:1977sc,Yanagida:1979as,PhysRevD.20.2986,GellMann:1980vs} and the baryon asymmetry of the universe via sphaleron-assisted conversion of the lepton asymmetry seeded by the out-of-equilibrium decay of the heavy right-handed neutrinos~\cite{Fukugita:1986hr}. Non-trivial fact resulting from the introduction of the three RH neutrinos is that the mixed anomaly of $[U(1)_{\rm B-L}]^{3}$ vanishes, which opens up the possibility that $U(1)_{\rm B-L}$ is a gauge symmetry of the theory. 

Motivated by these, we take $U(1)_{\rm B-L}$ as one of gauge symmetries of the theory from here on with the three heavy RH neutrinos. Then, what aspect of $U(1)_{\rm B-L}$ gauge theory can be invoked to improve the quality of axion?

As for the extension of the MSSM by $U(1)_{\rm B-L}$ gauge symmetry and $U(1)_{\rm PQ}$ global symmetry anomalous with respect to $SU(3)_{c}$, one may wonder if it is necessary to have the two sectors, i.e. matters carrying $U(1)_{\rm B-L}$ charges and $U(1)_{\rm PQ}$ charges, completely separated. Namely, we can ask if matters can carry both of charges simultaneously, still maintaining the original properties of each symmetry to resolve the original motivations including neutrino mass, leptogenesis and the strong CP problem. 

The axion quality problem can be the reason of the curiosity for the possibility of having matter fields bi-charged under $U(1)_{\rm B-L}$ and $U(1)_{\rm PQ}$. Carrying non-zero lepton numbers, three RH neutrinos are chiral in that they have the Majorana mass terms in the Lagrangian. Therefore, $U(1)_{\rm B-L}$ gauge invariance demands that the Majorana masses $M_{R}$ be understood as the spurion of $U(1)_{\rm B-L}$. In other words, $M_{R}$ arises from a vacuum expectation value (VEV) of a scalar $\Phi$ which causes the spontaneous breaking of $U(1)_{\rm B-L}$. 

Now let us denote for the moment the scalar inducing the spontaneous breaking of $U(1)_{\rm PQ}$ by $\Phi'$. Provided that both $\Phi$ and $\Phi'$ are bi-charged under $U(1)_{\rm B-L}$ and $U(1)_{\rm PQ}$, this can be of great help for alleviating the axion quality problem.\footnote{When the bi-charged scalars are considered, some of fermions coupled to them should be also properly bi-charged for the invariance of $U(1)_{\rm B-L}$ and $U(1)_{\rm PQ}$ at least at the renormalizable level.} $U(1)_{\rm B-L}$ gauge invariance does not allow for all the non-hermitian higher dimensional operators consisting of both $\Phi$ and $\Phi'$ except for those respecting $U(1)_{\rm B-L}$ gauge symmetry. Thereby the number of unwanted $U(1)_{\rm PQ}$-violating higher dimensional operators drastically decreases and the problem reduces to suppression of only few remaining $U(1)_{\rm B-L}$-invariant higher dimensional operators.

Let us illustrate the point with a concrete example of the charge assignment. For simplicity, let us assume $Q_{\rm B-L}[\Phi]=p$ and $Q_{\rm B-L}[\Phi']=-q$, ($p,q>0$), and $p$ and $q$ do not have any common divisor. Then, irrespective of $U(1)_{\rm PQ}$ charge assignment for the two scalars, only operators of the following type respect $U(1)_{\rm B-L}$ and thus can appear in the superpotential\footnote{Of course only those operators in Eq.~(\ref{eq:WPQviolating}) respecting $R$-symmetry can appear in the superpotential. But for illustration, let us only focus on the holomorphicity of operators as a condition to appear in the superpotential.}
\be
W\supset\sum_{n=1} M_{P}^{3}\left(\frac{\Phi^{q}\Phi'^{p}}{M_{P}^{p+q}}\right)^{n}\equiv\sum_{n=1}\mathcal{O}_{\cancel{PQ}}^{(n)} \,,
\label{eq:WPQviolating}
\ee
where the sum is over positive integers. This means, in the context of the axion quality problem, one only needs to make it sure that $\Delta V(\theta+\delta)$ contributed by these operators are small enough to guarantee $\Delta\theta_{\rm min}<10^{-10}$ for a given set of VEVs ($\langle\Phi\rangle,\langle\Phi'\rangle$)~\cite{Fukuda:2017ylt,Fukuda:2018oco,Choi:2020vgb}.

Now if we want to take advantage of the strategy presented above, then the $U(1)_{\rm B-L}$ gauge symmetry should be understood as the consequence of a linear combination of two $U(1)$ symmetries. And the other linear combination of the two $U(1)$s is identified with $U(1)_{\rm PQ}$. Suppose one is given two $U(1)$s, say $U(1)_{1}$ and $U(1)_{2}$, anomalous with respect to $SU(3)_{c}$ with different anomaly coefficients $Q_{1}$ and $Q_{2}$, and Noether currents $j_{1}^{\mu}$ and $j_{2}^{\mu}$ respectively, i.e.
\be
\partial_{\mu}j^{\mu}_{1}=Q_{1}\frac{g_{c}^{2}}{32\pi^{2}}G_{\mu\nu}^{a}\tilde{G}^{a\mu\nu}\quad,\quad\partial_{\mu}j^{\mu}_{2}=Q_{2}\frac{g_{c}^{2}}{32\pi^{2}}G_{\mu\nu}^{a}\tilde{G}^{a\mu\nu}\,.
\label{eq:anomalyfreetwoU(1)}
\ee
At this moment, $\Phi$ ($\Phi'$) is assumed to be charged only under $U(1)_{1}$ ($U(1)_{2}$).

It is always possible to find a linear combination $j_{AF}^{\mu}$ of $j_{1}^{\mu}$ and $j_{2}^{\mu}$ satisfying $\partial_{\mu}j_{AF}^{\mu}=0$ while an independent linear combination $j_{A}^{\mu}$ remains anomalous. As the continuous symmetry completely anomaly free at the quantum level, the $U(1)_{AF}$ symmetry generated by $j_{AF}^{\mu}$ can be taken as the gauge symmetry. In contrast, the other $U(1)_{A}$ generated by $j_{A}^{\mu}$ can be used to address the strong CP problem as NGB arising from the breaking of $U(1)_{A}$ can serve as the QCD axion. After reorganization for transition from the basis ($j_{1}^{\mu},j_{2}^{\mu}$) to ($j_{AF}^{\mu},j_{A}^{\mu}$), both $\Phi$ and $\Phi'$ become bi-charged under $U(1)_{AF}$ and $U(1)_{A}$.

In this work, we interpret $U(1)_{\rm B-L}$ gauge symmetry and $U(1)_{\rm PQ}$ as originated from the mechanism explained above. We identify $U(1)_{\rm B-L}$ gauge symmetry and $U(1)_{\rm PQ}$ global symmetry with $U(1)_{AF}$ and $U(1)_{A}$. This implies that matter fields of the two sectors may be bi-charged under $U(1)_{\rm B-L}$ gauge symmetry and $U(1)_{\rm PQ}$. For a given model, starting from renormalizable Lagrangian respecting $U(1)_{\rm B-L}$, we can infer $U(1)_{\rm PQ}$-charge assignment. From this, we can get information for what $p$, $q$ are and the most dangerous $n$'s. We will go through this procedure in Sec.~\ref{sec:model}.

After the spontaneous breaking of $U(1)_{\rm B-L}$ and $U(1)_{\rm PQ}$, by comparison of (1) the kinetic terms of $\Phi=(f/\sqrt{2})e^{i(A/f)}$ and $\Phi'=(f'/\sqrt{2})e^{i(A'/f')}$ with $\langle\Phi\rangle=f/\sqrt{2}$ and $\langle\Phi'\rangle=f'/\sqrt{2}$ and (2) that of the axion field $a$ and $U(1)_{AF}$ gauge boson's mass term, one can obtain the expression for the axion $a$ as a linear combination of $A$ and $A'$, and that for the axion decay constant $F_{a}$~\cite{Fukuda:2017ylt}
\be
\begin{bmatrix}
           a \\
           b 
         \end{bmatrix}=\frac{1}{\sqrt{p^{2}f^{2}+q^{2}f'^{2}}}\begin{bmatrix}
     -qf' & -pf \\
     pf & -qf' 
\end{bmatrix} \begin{bmatrix}
           A \\
           A' 
         \end{bmatrix}\quad,\quad F_{a}=\frac{ff'}{\sqrt{p^{2}f^{2}+q^{2}f'^{2}}}
         \label{eq:expressionofaxion}
\ee
where $b$ is the NGB eaten by $U(1)_{AF}$ gauge boson.\footnote{From the expression of the axion in Eq.~(\ref{eq:expressionofaxion}), one can obtain $a/F_{a}=-q(A/f)-p(A'/f')$ which must be invariant under the gauge transformation of $U(1)_{\rm B-L}$, i.e. $A/f\rightarrow A/f + p\alpha_{\rm B-L}$ and $A'/f'\rightarrow A'/f' - q\alpha_{\rm B-L}$ with $\alpha_{\rm B-L}$ the unit of the phase rotation under $U(1)_{\rm B-L}$ transformation. Identification of $a/F_{a}=-q(A/f)-p(A'/f')$ with Eq.~(\ref{eq:expressionofaxion}) gives us the expression of $F_{a}$ in terms of $f$ and $f'$.}

\subsection{Gauged $R$-symmetry}
\label{sec:m32}

In Sec.~\ref{sec:twoU(1)s}, we discussed how the global $U(1)_{\rm PQ}$ symmetry can be protected from $U(1)_{\rm PQ}$-violating nonrenormalizable operators with the aid of the gauged $U(1)_{B-L}$ symmetry, excepting those in Eq.~(\ref{eq:WPQviolating}). With the hope to make the mechanism complete on its own, one may try to choose a set of ($p,-q$) such that $p+q$ is large enough to make $\Delta V(\theta+\delta)$ contributed by the operator $\sim\Phi^{p}\Phi'^{q}$ in Eq.~(\ref{eq:WPQviolating}) with $n=1$ produce $\Delta\bar{\theta}_{\rm min}<10^{-10}$ for a given $(\langle\Phi\rangle,\langle\Phi'\rangle)$~\cite{Fukuda:2017ylt,Fukuda:2018oco}. Such a selection of $p$ and $q$, however, is completely lack of any underlying physics and rule, and thus we see that it is not logical enough to serve as a solution to the axion quality problem.

Instead, in this section, we discuss the use of a gauged $R$-symmetry to suppress the operators in Eq.~(\ref{eq:WPQviolating}). Insofar as a theory is embedded in a SUGRA framework, there always exists a $R$-symmetry and thus every operator appearing in the superpotential is required to respect $R$-symmetry. This requirement can provide us with a powerful way of handling the unwanted operators in Eq.~(\ref{eq:WPQviolating}) provided there is a logical and systematic way of assigning $R$-charges to $\Phi$ and $\Phi'$ in a model. In this section, we focus on the role of $R$-symmetry to suppress remaining unwanted operators in Eq.~(\ref{eq:WPQviolating}) and $\Delta V(\theta+\delta)$ in SUGRA. Then in Sec.~\ref{sec:model}, we discuss in detail how the model of our interest can lead to a small enough $\Delta\bar{\theta}_{\rm min}<10^{-10}$ in accordance with a systematic $R$-charge assignment. 

\subsubsection{$R$-symmetry-assisted suppression of $\Delta V(\theta+\delta)$}
\label{sec:suppress}
As a matter of fact, even if operators in Eq.~(\ref{eq:WPQviolating}) respect $U(1)_{\rm B-L}$, they need to be modified depending on $R[\Phi]$ and $R[\Phi']$, and a type of $R$-symmetry in order to appear in the superpotential. Envisioning a model in $\mathcal{N}=1$ SUGRA, when $U(1)_{R}$ is assumed, only operators with $R$-charge 2 can appear in the superpotential. Instead if a discrete R-symmetry $Z_{NR}$ ($N\in \mathbb{N}$\,\,\&\,\,$N>2$) is assumed, operators themselves or those multiplied by some powers of $m_{3/2}$ must carry $R$-charges 2 modulo $N$ in order to appear in the superpotential.\footnote{Here $m_{3/2}\equiv|F_{Z}|/(\sqrt{3}M_{P})$ is a gravitino mass with $F_{Z}$ the auxiliary field component of a SUSY-breaking field $Z$. With SUSY-breaking, the vanishingly small cosmological constant requires a constant term in the superpotential satisfying $W_{0}=m_{3/2}M_{P}^{2}$. Because of this, we have $R[m_{3/2}]=2$}

Now the interesting case is when $|R[\mathcal{O}_{\cancel{PQ}}^{(n=1)}]|>2$ holds for $\mathcal{O}_{\cancel{PQ}}^{(n=1)}$ in Eq.~(\ref{eq:WPQviolating}). In this case, under $U(1)_{R}$, all the operators in Eq.~(\ref{eq:WPQviolating}) are not allowed since their $R$-charges can never be $2$. Therefore, if this ideal situation can be consistently realized in a model, it will provide the complete solution to the axion quality problem. 

On the other hand, under a $Z_{NR}$, operators must either be multiplied by some powers of $m_{3/2}/M_{P}$ to have $R$-charge 2 modulo $N$ and to appear in $W$ or disappear from $W$. Given that the scalar potential $V$ in SUGRA is given by 
\be
V=e^{K/M_{P}^{2}}\left[\sum_{a,b}\left(\frac{\partial^{2}K}{\partial \Theta_{a}\partial \Theta_{b}^{*}}\right)^{-1}D_{\Theta_{a}}WD_{\Theta_{b}^{*}}W^{*}-3e^{K/M_{P}^{2}}\frac{|W|^{2}}{M_{P}^{2}}\right]\,,
\label{eq:VSUGRA}
\ee
where $D_{\Theta_{a}}W=(\partial W/\partial \Theta_{a})+(W/M_{P}^{2})(\partial K/\partial \Theta_{a})$ and $\Theta_{a}$ is a chiral superfield, in the end operators are to be multiplied by a single $m_{3/2}^{\dagger}$ further to appear in $V$. In this case, if $m_{3/2}$ is sufficiently small as compared to $M_{P}$, a large suppression of $\Delta V(\theta+\delta)$ and thus $\Delta\bar{\theta}_{\rm min}$ can be induced. In \cite{Choi:2020vgb}, by using 
$m_{3/2}=\mathcal{O}(1){\rm eV}$, this strategy was taken to have the operator with $n=1$ in Eq.~(\ref{eq:WPQviolating}) sufficiently suppressed.

\subsubsection{$U(1)_{R}$ or $Z_{NR}$?}
\label{sec:suppress}
As was pointed out in the previous section, depending on if a discrete $R$-symmetry in low energy is a remnant of the spontaneous breaking of a gauged $U(1)_{R}$ or not, understanding for the contribution to $\Delta V(\theta+\delta)$ from operators in Eq.~(\ref{eq:WPQviolating}) can be varied. 

If a gauged $Z_{NR}$ is to be understood as the remnant of a broken gauged $U(1)_{R}$, the mixed anomaly free conditions for $U(1)_{R}$ and other gauge groups must be satisfied. This can be achieved by either an appropriate $R$-charge assignment of particle contents of the theory or the help of the Green-Schwarz (GS) mechanism~\cite{Green:1984sg}. The former case was actually investigated in \cite{Chamseddine:1995gb} and for a generation-independent $R$-charge assignment, it was shown that $U(1)_{R}$ extension of MSSM remains anomalous unless an extra $SU(3)_{c}$ color octet is introduced. Another investigation was made in \cite{Castano:1995ci} with the additional chiral superfields including color-triplet Higges  SUSY-breaking fields to the MSSM and no rational $R$-charge assignment was found for $R$-anomaly cancellation.

Alternatively one can rely on the $R$-anomaly cancellation with the aid of the GS mechanism provided
\be
\frac{\mathcal{A}_{1}}{k_{1}}=\frac{\mathcal{A}_{2}}{k_{2}}=\frac{\mathcal{A}_{3}}{k_{3}}=\frac{\mathcal{A}_{B-L}}{k_{B-L}}
\label{eq:GS}
\ee
where $k_{1}$, $k_{2}$, $k_{3}$ and $k_{B-L}$ are Ka$\check{c}$-Moody levels of $U(1)_{Y}$, $SU(2)_{L}$, $SU(3)_{c}$ and $U(1)_{B-L}$, and $\mathcal{A}_{i}'$s are the mixed anomalies of $U(1)_{R}-[G]^{2}$ with $G$ each of four gauge groups. The problem is that the normalization for charges under $U(1)$ symmetries, and $k_{1}$ and $k_{B-L}$ are uncertain so that one cannot be sure of equalities in Eq.~(\ref{eq:GS}) unless the whole theory is known.\footnote{Even if we envision the gauge coupling unification among the SM gauge group, still $k_{B-L}$ still remains uncertain. Another challenge is to make it sure that other anomalies including $U(1)_{R}^{2}-U(1)_{Y}$, $U(1)_{R}^{2}-U(1)_{B-L}$, $U(1)_{R}^{3}$ and $U(1)_{R}-[{\rm gravity}]^{2}$ vanish as well.}

Given the practical difficulties encountered in satisfying $R$-anomaly free conditions, in this work we restrict ourselves to the case where $Z_{NR}$ is the $R$-symmetry that the operators in Eq.~(\ref{eq:WPQviolating}) should respect. In this case, along with the interactions we have in the superpotential, very useful constraint on $R$-charge assignment of massless fermions is provided by the anomaly free conditions of mixed anomalies $Z_{NR}-[SU(2)_{L}]^{2}$ and $Z_{NR}-[SU(3)_{c}]^{2}$. Note that given a discrete symmetry $Z_{N}$, the contribution to the mixed anomaly $Z_{N}-[SU(N)]^{2}$ from the massless and massive fermions acquiring mass from the breaking $U(1)\rightarrow Z_{N}$ is separately cancelled and thus the anomaly free condition for $Z_{N}-[SU(N)]^{2}$ is insensitive to heavy particle spectrum~\cite{Ibanez:1991hv,Ibanez:1992ji}.\footnote{Differing from $Z_{N}-[SU(N)]^{2}$, the mixed cubic anomaly $Z_{N}^{3}$ or the gravitational anomaly do not give a useful constraint on the massless fermions. This is because the required anomaly free conditions for these mixed anomalies heavily depend on whether the heavy particles obtaining masses from the breaking $U(1)\rightarrow Z_{N}$ are Dirac or Majorana fermions~\cite{Ibanez:1991hv,Ibanez:1992ji}. The mixed cubic anomaly $Z_{N}^{3}$ is affected by both Dirac and Majorana massive fermions and their charge assignment while $Z_{N}$-gravitational anomaly may be contributed by massive Majorana fermions for cancellation. Essentially for these mixed anomalies, the contribution to the anomaly from the massless and massive fermions are not decoupled.} 

The anomaly free condition for the mixed anomaly $Z_{NR}-[SU(M)]^{2}$ reads
\be
\mathcal{A}_{NRM}\equiv2T({\bf{\rm Adj}})+\sum_{i}2T(R_{i})\times(R[\Phi_{i}]-1)=0\quad{\rm mod}\quad M\,,
\label{eq:mixedanomalyfree}
\ee
where $T(R)$ is the Dynkin index for the representation $R$ of $SU(M)$ and the sum runs over different matter fields. Given Eq.~(\ref{eq:mixedanomalyfree}), now we realize that there can be two options: either $\mathcal{A}_{NRM}=Mk\neq0$ with $k\,\,({\rm or}\,-k)\in\mathbb{N}$ or $\mathcal{A}_{NRM}=0$. 

We recall that for the purpose of addressing the axion quality problem, of course a larger $N$ for $Z_{NR}$ is better. For the first option, we may have difficulty in taking a large enough $N$ for $Z_{NR}$ {\it unless we either assign intentionally large 
$R$-charges to fields contributing to $\mathcal{A}_{NRM}$ or introducing many fields carrying $R$-charge that contribute to $\mathcal{A}_{NRM}\neq0$.} In contrast, intriguingly the second option gives us the logically supported full freedom for the choice of any $N$ for $Z_{NR}$ without the weird set-up in the hidden sector.

Therefore, in the coming model building part in Sec.~\ref{sec:app}, we consider the case of $\mathcal{A}_{NRM}=0$. Although we assume $Z_{NR}$ in the theory, because of the freedom in choosing $Z_{NR}$, the theory can benefit from the power to control operators in Eq.~(\ref{eq:WPQviolating}) as strong as $U(1)_{R}$. In Sec.~\ref{sec:model}, $\mathcal{A}_{NRM}=0$ will become the guiding principle to specify $R[\Phi]$ and $R[\Phi']$. And this will be directly related to the prediction of the theory for the axion quality.

We conclude this section by emphasizing the big difference between a gauged discrete symmetry $Z_{N}$ and the gauged discrete $R$-symmetry $Z_{NR}$. Naively one may guess that when PQ scalar carries the charge of $Z_{N}$, imposing a gauged discrete $Z_{N}$ with a large $N$ can resolve the axion quality problem very easily because its presence is expected to suppress most of the higher dimensional operators consisting of the PQ scalar. 

However, we need to remember that the gauged $Z_{N}$ should be subject to the mixed anomaly free conditions with $G_{\rm SM}$. Already within the MSSM, it is readily possible to have the mixed anomalies of $Z_{N}$ with non-Abelian gauge groups in MSSM equal to 0 mod $N$ (see Appendix~\ref{sec:appendixA}). Thus it is not necessary to extend the MSSM matter sector for gauging $Z_{N}$. On the other hand, apparently colored fermions coupled to the PQ scalar should carry a charge of $Z_{N}$ and therefore we are aware of at least these fermions' non-zero contribution to the mixed anomaly $Z_{N}-[SU(3)]^{2}$. Since the MSSM itself contributes to the mixed anomaly by 0 mod $N$, new colored fermion's contribution itself must be an integer multiple of $N$. This means that there should be at least $N$ different species of colored fermions as the Yukawa couplings of the PQ scalar to these colored fermions carry $Z_{N}$-charge $N$.\footnote{The introduction of $N$ species of colored fermions can be avoided if one considers the possibility of the mixed anomaly cancellation via Green-Schwarz mechanism~\cite{Babu:2002ic}. However, given the uncertainty of $k_{1}$ and $k_{\rm B-L}$, it is not guaranteed whether a model can really satisfy Eq.~(\ref{eq:GS}).} Therefore, using the gauged $Z_{N}$ with a large $N$ for axion quality problem seems to be in need of corresponding a large number of new colored fermions as the price to pay.

On the contrary, the discrete $R$-symmetry is intrinsically different: new colored fermion can contribute $\mathcal{A}_{NRM}\neq0\,\,{\rm mod}\,\,N$ to the mixed anomalies, providing the logical reason to avoid to introduce many new fermions. Recall that already there is a unavoidable non-vanishing contribution to $\mathcal{A}_{NR2}\neq0\,\,{\rm mod}\,\,N$ and $\mathcal{A}_{NR3}\neq0\,\,{\rm mod}\,\,N$ from the MSSM particle contents (see Appendix~\ref{sec:appendixB}).\footnote{For the high quality axion, we have to consider $N>8$, which will be explained in Sec.~\ref{sec:axionquality}. So $\mathcal{A}_{NR2}=0\,\,{\rm mod}\,\,N$ and $\mathcal{A}_{NR3}=0\,\,{\rm mod}\,\,N$ cannot be the case, which is possible for $N=6$.} This means that newly added, but few colored fermions coupled to the PQ scalar can cancel this existing MSSM contribution to make $\mathcal{A}_{NR2}=0$ and $\mathcal{A}_{NR3}=0$. Hence however a large $N$ one may imagine for $Z_{NR}$, it never requires the corresponding a huge number of new colored fermions charged under $Z_{NR}$ because the new fermion's contribution itself needs not be 0 mod $N$. This crucial difference between $Z_{N}$ and $Z_{NR}$ determines whether an arbitrary choice of a large enough $N$ for addressing the axion quality problem can be logically justified or not.

\section{High Quality Axion}
\label{sec:app}
\setcounter{equation}{0}

In this section, based on the two symmetries that we introduced in Sec.~\ref{sec:twosym}, we establish a concrete model with additional fields on top of MSSM particle contents with the purpose of addressing the axion quality problem. As will be shown, $U(1)_{\rm PQ}$ global symmetry emerges as an accidental symmetry of the hidden sector which is bi-charged under $U(1)_{\rm B-L}$ and $U(1)_{\rm PQ}$. The Planck-suppressed $U(1)_{\rm PQ}$-violating operators will be shown to be naturally suppressed thanks to the gauged $U(1)_{\rm B-L}$ and $Z_{NR}$. With the additional fields, renormalization group evolution (RGE) of $SU(3)_{c}$ gauge coupling is modified in the energy regime above $U(1)_{\rm B-L}$ and $U(1)_{\rm PQ}$ breaking scale. We shall discuss if this causes another dangerous $\Delta V(\theta+\delta)$ and show that the corresponding $\Delta\bar{\theta}_{\rm min}$ does not exceed $10^{-10}$ as long as a SUSY-breaking scale is below $\sim10^{14}{\rm GeV}$.

\subsection{Model}
\label{sec:model}
As the symmetry group of the model, we consider
\be
 G_{\rm sym}=\underbrace{G_{\rm SM}\otimes U(1)_{\rm B-L}\otimes Z_{NR}}_{\rm gauge}\otimes\underbrace{U(1)_{\rm PQ}}_{\rm global}\,,
\label{eq:symmetry}
\ee
where $G_{\rm SM}=SU(3)_{c}\otimes SU(2)_{L}\otimes U(1)_{Y}$ is the MSSM gauge group and $N$ in the discrete $R$-symmetry is unspecified at the moment. As discussed in Sec.~\ref{sec:twoU(1)s}, we assume the presence of two $U(1)$s anomalous with respect to $SU(3)_{c}$ and we interpret the usual $U(1)_{\rm B-L}$ gauge theory with the three RH neutrinos extending the SM as the anomaly free linear combination of two $U(1)$s anomalous with respect to $SU(3)_{c}$. The other independent anomalous linear combination is identified as the global $U(1)_{\rm PQ}$. 

\begin{table*}[h]
\centering
\begin{tabular}{|c||c|c|c|c|c|c|c|c|c|c|c|} \hline
 & $\bf{5}^{*}_{i}$ & $\bf 10_i$ & $N_i$ & $H_u$ & $H_d$ \\
\hline
$U(1)_{B-L}$ & -3& 1& 5& -2& 2\\
\hline
$Z_{NR}$ & 0 & 0 & 0 & 2 & 2 \\
\hline
\end{tabular}
\caption{Charge assignment of MSSM particle contents.}
\label{table:qn} 
\end{table*}

We first discuss the anomaly free conditions of $U(1)_{\rm B-L}$ and additional matter contents. As the gauged symmetry, one needs to make it sure that the mixed anomalies $U(1)_{\rm B-L}-G_{\rm SM}^{2}$, $U(1)_{\rm B-L}^{3}$ and $U(1)_{\rm B-L}-[{\rm Gravity}]^{2}$ vanish. It is well known that the first one readily vanishes within the MSSM, and the second and third one can also vanish if those are further contributed by the three RH neutrinos $N_{i}$ ($i=1,2,3$). In Table.~\ref{table:qn}, we show the charge assignment of MSSM fields under $U(1)_{\rm B-L}$ and $Z_{NR}$ which respects the following Yukawa couplings in the MSSM and the Higgsino mass term\footnote{Our model differes from the usual $SU(5)$ GUT model in that the particle contents do not contain the colored Higgs triplet. Later we will show that the condition $\mathcal{A}_{NR3}=\mathcal{A}_{NR2}=0$ achieved in the model is crucial in justifying the choice of an arbitrarily large $N$ in $Z_{NR}$. For this purpose, we do not introduce the colored Higgs triplet, which will spoil $\mathcal{A}_{NR3}=\mathcal{A}_{NR2}=-6$ within MSSM. When $\mathcal{A}_{NR3}\neq\mathcal{A}_{NR2}$ is the case before introducing extra matter fields, it becomes difficult to make the part of the mixed anomalies of $U(1)_{\rm B-L}$, i.e. $[U(1)_{\rm B-L}]^{3}$ and $[U(1)_{\rm B-L}]-[{\rm gravity}]^{2}$ vanish.}
\be
W\supset y_{u,ij}{\bf 10}_{i}{\bf 10}_{j}H_{u}+y_{d,ij}{\bf 10}_{i}{\bf 5}^{*}_{j}H_{d}+y_{\nu,ij}N_{i}{\bf 5}^{*}_{j}H_{u}+M_{R,i}N_{i}N_{i}+\mu H_{u}H_{d}\,.
\label{eq:MSSMW}
\ee
One can indeed see that the three anomaly free conditions for $U(1)_{\rm B-L}$ are satisfied. Note that the RH neutrino mass $M_{R,i}$ serves as the spurion field with $R[M_{R,i}]=2$ which is originated from condensation of $\bar{\Phi}$ in Table.~\ref{table:qn2}.

As for the R-charge assignment, we notice that there are five conditions to impose for determining charges of five matter fields in Table.~\ref{table:qn}: Four Yukawa couplings in Eq.~(\ref{eq:MSSMW}) give four conditions and the other last condition comes from the mixed anomaly condition of $R$-symmetry within MSSM. Here, the Majorana mass term is understood to arise from a Yukawa interaction. As was discussed in Sec.~\ref{sec:suppress}, the mixed anomalies $\mathcal{A}_{NR2}$ and $\mathcal{A}_{NR3}$ are subject to the anomaly free condition within the MSSM since these must be insensitive to a heavy fermion contribution. Therefore, the following solid argument for $\mathcal{A}_{NR2}$ and $\mathcal{A}_{NR3}$ can be made (see Appendix.~\ref{sec:appendixB})
\be
\mathcal{A}_{NR2}-\mathcal{A}_{NR3}=0\,\,{\rm mod}\,\,N\,. 
\label{eq:anomalyfree}
\ee
In computing $\mathcal{A}_{NR2}$ and $\mathcal{A}_{NR3}$ based on Eq.~(\ref{eq:mixedanomalyfree}), we encounter the condition~\cite{Harigaya:2013vja} 
\be
R[H_{u}]+R[H_{d}]=4\,\,{\rm mod}\,\,N\,.
\ee
Thus, along with the last condition in Eq.~(\ref{eq:anomalyfree}), the four Yukawa determines $R$-charges of matter fields in Table.~\ref{table:qn} completely. Later we will set $R[M_{R}]=2$, which determines R[N] and the rest of $R$-charges of MSSM matter sector as shown in Table.~\ref{table:qn}.

For $U(1)_{\rm PQ}$ to be anomalous with respect to $SU(3)_{c}$, the model needs at least one colored matter field charged under $U(1)_{\rm PQ}$. As $\Phi$ and $\Phi'$ are bi-charged, the matter fields should be bi-charged as well for forming the Yukawa couplings with $\Phi$ and $\Phi'$. Such an addition will make a new contribution to all the mixed anomalies $U(1)_{\rm B-L}-G_{\rm SM}^{2}$, $U(1)_{\rm B-L}^{3}$ and $U(1)_{\rm B-L}-[{\rm Gravity}]^{2}$ and thus the added new fields should make the net zero contribution to these anomalies.

On the other hand, if there are no additional fields carrying isospin that accompany the introduction of the additional colored fields, the mixed anomalies $Z_{NR}-[SU(3)]^{2}$ and $Z_{NR}-[SU(2)]^{2}$ can deviate from each other. This is because the contribution to the mixed anomalies $Z_{NR}-[SU(3)]^{2}$ and $Z_{NR}-[SU(2)]^{2}$ within MSSM is identically $-6$. The deviation is problematic for having the discrete gauged $Z_{NR}$ symmetry. So whatever new additional field charged under $G_{\rm SM}$ is introduced, as the minimum requirement for having $Z_{NR}$ gauged anomaly free symmetry, there should be identical changes in $\mathcal{A}_{NR3}$ and $\mathcal{A}_{NR2}$ (see Eq.~(\ref{eq:mixedanomalyfree}) for computation of these coefficients). Also we keep in mind that as explained in Sec.~\ref{sec:suppress}, for better addressing the axion quality problem, we want to have the corresponding anomaly coefficients fulfill the condition $\mathcal{A}_{NR3}=\mathcal{A}_{NR2}=0$.

With that being said, we introduce the set (${\bf 5}, {\bf 5}^{*}$) bi-charged under $U(1)_{\rm B-L}$ and $U(1)_{\rm PQ}$ for inducing the coupling given in Eq.~(\ref{eq:anomalyterm}) and also for an identical change in the mixed anomalies $Z_{NR}-[SU(3)]^{2}$ and $Z_{NR}-[SU(2)]^{2}$ per a new bi-charged matter field. Note that we introduce the set to have the gauge invariance of operators containing the new fields .\footnote{We avoid the coupling of the new fields to the MSSM matter fields as that restricts charges of $U(1)$s carried by the new fields.}  Then, how many sets of (${\bf 5}, {\bf 5}^{*}$) should we introduce to keep $U(1)_{\rm B-L}$ anomaly free?

Note that ${\bf 5}^{*}$ is just the counterpart of ${\bf 5}$ for forming a gauge invariant Yukawa coupling with $\Phi$ and $\Phi'$. And it cannot be the case that $Q_{\rm B-L}[{\bf 5}^{*}]=-Q_{\rm B-L}[{\bf 5}]$ since there cannot be Yukawa coupling to $\Phi$ or $\Phi'$ with such charges. This implies that we need at least more than one set of (${\bf 5}, {\bf 5}^{*}$). Thus this question is equivalent to asking the minimum number of new ${\bf 5}$'s which carry distinct $Q_{\rm B-L}$ and render the mixed anomalies $U(1)_{\rm B-L}^{3}$ and 
$U(1)_{\rm B-L}-[{\rm Gravity}]^{2}$ vanish. 

The minimum number of the necessary sets of (${\bf 5}, {\bf 5}^{*}$) turns out to be five~\cite{Batell:2010bp,Nakayama:2011dj,Costa:2020dph}.\footnote{Two and four sets of (${\bf 5}, {\bf 5}^{*}$) can easily lead to the gauge invariant ${\bf 5}{\bf 5}^{*}$s forming Dirac mass terms without coupling to $\Phi$ and $\Phi'$. So it is out of interest. Three sets of (${\bf 5}, {\bf 5}^{*}$) cannot solve the anomaly free conditions for $U(1)_{\rm B-L}^{3}$ and 
$U(1)_{\rm B-L}-[{\rm Gravity}]^{2}$. For application to the other phenomenologies, see also \cite{Nakayama:2018yvj,Choi:2020tqp,Choi:2020udy,Choi:2020nan}.} We show the charge assignment of the five sets of (${\bf 5}, {\bf 5}^{*}$) and chiral superfields for breaking two $U(1)$s in Table.~\ref{table:qn2}.\footnote{One may wonder how the cosmology is affected by the presence of $\Psi$'s and $\bar{\Psi}$'s. There can be two cases depending on how the reheating temperature $T_{\rm RH}$ is compared to $v$. If $T_{\rm RH}>v$ holds, we checked that $\Psi$'s and $\bar{\Psi}$'s are thermalized by the MSSM thermal bath at the reheating era via the MSSM particle scattering mediated by $B-L$ gauge boson. And once $\Psi$'s and $\bar{\Psi}$'s become non-relativistic, they are simply Boltzmann suppressed and integrated-out. In contrast, if $T_{\rm RH}<v$ holds, $\Psi$'s and $\bar{\Psi}$'s do not have any chance to be produced in the MSSM thermal bath as their production is kinematically suppressed. Note that the inflaton is assumed to be neutral to $U(1)_{\rm B-L}$ for the successful slow roll inflation. So $\Psi$'s and $\bar{\Psi}$'s production via the inflaton decay is prohibited as well. Thus the presence of $\Psi$'s and $\bar{\Psi}$'s do not cause any danger in cosmology.} As one can check, the quantum numbers in Table.~\ref{table:qn2} accomplish (1) $\mathcal{A}_{NR3}=\mathcal{A}_{NR2}=0$ (2) zero contribution to the mixed anomalies $U(1)_{\rm B-L}^{3}$ and 
$U(1)_{\rm B-L}-[{\rm Gravity}]^{2}$ and (3) the anomalous $U(1)_{\rm PQ}$. As for $Q_{\rm PQ}$-assignment, one can impose nonzero $Q_{\rm PQ}$ to $(\Phi,\overline{\Phi})$ and $\Psi_{Q_{\rm B-L}}$s with $Q_{\rm B-L}=-1,-5,-9$ properly. Here we don't do that just for keeping the minimality of the model. Also the normalization of $Q_{\rm PQ}$ and $Q_{\rm B-L}$ can be varied, which does not spoil the anomaly free conditions.

\begin{table*}[h]
\centering
\begin{tabular}{|c||c|c|c|c|c|c|c|c|c|} \hline
 & $\Phi$ & $\overline{\Phi}$ & $\Phi'$ & $\overline{\Phi}'$ & $X$ & $Y$ \\
\hline
$U(1)_{B-L}$ & 10& -10& -15& 15& 0& 0\\
\hline
$U(1)_{\rm PQ}$ & 0 & 0 & 1 & -1 & 0 & 0 \\
\hline
$Z_{NR}$ & -2 & 2 & 0 & 0 & 2 & 2 \\
\hline
\end{tabular}

\vskip 0.2in

\begin{tabular}{|c||c|c|c|c|c|c|c|c|c|c|c|c|c|} \hline
 & $\Psi_{-1}$ & $\Psi_{-5}$ & $\Psi_{-9}$ & $\Psi_{7}$ & $\Psi_{8}$ & $\overline{\Psi}_{-1}$ &  $\overline{\Psi}_{-5}$ &  $\overline{\Psi}_{-9}$ & $\overline{\Psi}_{7}$& $\overline{\Psi}_{8}$ \\
\hline
$U(1)_{B-L}$ & -1& -5& -9& 7& 8& -1& -5& -9 & 7 &8\\
\hline
$U(1)_{\rm PQ}$ & 0 & 0 & 0 & 0 & 0 & 0& 0& 0& -1& -1 \\
\hline
$Z_{NR}$ & 2 & 2 & 2 & 1 & 1 & 2 & 2 & 2& 1&1 \\
\hline
$SU(5)$ & ${\bf 5}$ & ${\bf 5}$ & ${\bf 5}$ & ${\bf 5}$ & ${\bf 5}$ & ${\bf 5}^{*}$ & ${\bf 5}^{*}$ & ${\bf 5}^{*}$& ${\bf 5}^{*}$&${\bf 5}^{*}$ \\
\hline
\end{tabular}
\caption{Charge assignment of the newly introduced hidden sector.}
\label{table:qn2} 
\end{table*}

Based on the quantum numbers, now we can have the superpotential of the hidden sector
\be
W_{\rm hidden}\supset W_{\cancel{U(1)}}+W_{{\rm Yuk}}\,,
\label{eq:Whidden}
\ee
where $W_{\cancel{U(1)}}$ and $W_{{\rm Yuk}}$ are given by 
\be
W_{\cancel{U(1)}}=X(2\Phi\overline{\Phi}-v^{2})+Y(2\Phi'\overline{\Phi}'-v'^{2})\,,
\label{eq:Wu1}
\ee
and\footnote{As both $\bar{\Phi}$ and $\bar{\Phi}'$ do not couple to the matter fields and are singlet under $SU(2)$ and $SU(3)$ of MSSM, they can enjoy a separate discrete gauge symmetry. This can prevent the operators like $\mu^{2}\bar{\Phi}^{3}\bar{\Phi}^{'\,2}$ from contributing to the QCD axion potential.} 
\be
W_{\rm Yuk}\supset\Phi(\Psi_{-1}\overline{\Psi}_{-9}+\Psi_{-9}\overline{\Psi}_{-1})+\Phi\Psi_{-5}\overline{\Psi}_{-5}+\Phi'(\Psi_{7}\overline{\Psi}_{8}+\Psi_{8}\overline{\Psi}_{7})\,.
\label{eq:Wyuk}
\ee
In Eqs.~(\ref{eq:Wu1}) and (\ref{eq:Wyuk}), we omit the dimensionless coupling constants for simplicity. In this work we consider the case of $v\simeq v'$ as hierarchy among $v$ and $v'$ is irrelevant as far as the axion quality problem is concerned. 

After the spontaneous breaking of two $U(1)$s, we have $\Phi=(v/\sqrt{2})e^{A/v}$, $\overline{\Phi}=(v/\sqrt{2})e^{-A/v}$, $\Phi'=(v'/\sqrt{2})e^{A'/v'}$ and $\overline{\Phi}'=(v'/\sqrt{2})e^{-A'/v'}$ with $A$ and $A'$ the chiral superfields serving as Goldstone multiplets. From Eq.~(\ref{eq:expressionofaxion}), we can obtain the form of axion $a$ and the NGB of the broken $U(1)_{\rm B-L}$. In terms of $A$ and $A'$, the axion superfield $\mathcal{A}$ can be written as
\be
\frac{\mathcal{A}}{f_{a}}=(-3)\frac{{\rm Im}(A)}{v}-(2)\frac{{\rm Im}(A')}{v'},\qquad f_{a}=\frac{vv'}{\sqrt{(2)^{2}v^{2}+(-3)^{2}v'^{2}}}\,.
\label{eq:afa}
\ee
One can observe the invariance of the axion superfield under $U(1)_{\rm B-L}$ transformation, i.e. $\mathcal{A}\rightarrow \mathcal{A}+i(10\alpha_{\rm B-L})$ and $\mathcal{A}'\rightarrow \mathcal{A}'-i(15\alpha_{\rm B-L})$. We identify the axion $a$ with $a=\sqrt{2}{\rm Im}[\mathcal{A}]$ and the effective axion decay constant $F_{a}$ with $F_{a}=\sqrt{2}f_{a}$.

We conclude this section by commenting on the gauge coupling unification of the model. As the newly introduced matter fields in Table.~2 transform as fundamental and anti-fundamental representation of $SU(5)$, the gauge coupling unification that the MSSM features remains unaffected. We checked that the unification takes place at $\Lambda_{\rm GUT}\simeq1.8\times10^{16}{\rm GeV}$ with $\alpha_{\rm GUT}\simeq0.06$.

\subsection{Axion Quality}
\label{sec:axionquality}

For the model established in Sec.~\ref{sec:model}, now we are in a position to discuss the axion quality. Above all, the greatest worrisome for an axion quality is the operator with $n=1$ in Eq.~(\ref{eq:WPQviolating}). Given $Q_{\rm B-L}$ in Table.~\ref{table:qn2}, we can specify $(p,q)=(3,2)$. The operator of the greatest concern reads
\be
\mathcal{O}_{\cancel{PQ}}^{(n=1)}=c_{1}M_{P}^{3}\frac{\Phi^{3}\Phi'^{2}}{M_{P}^{5}}\,,
\label{eq:O1}
\ee
where $c_{1}$ is a dimensionless coefficient of the operator. Unless largely suppressed, this operator will cause a ridiculously large $\Delta\bar{\theta}_{\rm min}$. Then how does the model logically suppress it?

Notice that from the superpotential in Eq.~(\ref{eq:Wyuk}) preserving the $U(1)_{\rm B-L}$ gauge invariance, we obtain the following conditions for $R$-charge assignments of the hidden sector
\be
R[\Phi]=\frac{1}{3}(6-R[\Psi_{-1}\bar{\Psi}_{-9}]+R[\Psi_{-9}\bar{\Psi}_{-1}]+R[\Psi_{-5}\bar{\Psi}_{-5}])\quad,\quad R[\Phi']=\frac{1}{2}(4-R[\Psi_{7}\bar{\Psi}_{8}]+R[\Psi_{8}\bar{\Psi}_{7}])\,.
\label{eq:Rcharge1}
\ee
On the other hand, for having $\mathcal{A}_{NR3}=\mathcal{A}_{NR2}=0$ in Eq.~(\ref{eq:mixedanomalyfree}) when these are contributed by all the fields charged under $SU(3)_{c}$ and $SU(2)_{L}$, we need\footnote{Recall that the MSSM fields' contribution to $\mathcal{A}_{NR3}$ and $\mathcal{A}_{NR2}$ is given by $\mathcal{A}_{NR3}=\mathcal{A}_{NR2}=-6$.}
\be
R[\Psi_{-1}\bar{\Psi}_{-9}]+R[\Psi_{-9}\bar{\Psi}_{-1}]+R[\Psi_{-5}\bar{\Psi}_{-5}]+R[\Psi_{7}\bar{\Psi}_{8}]+R[\Psi_{8}\bar{\Psi}_{7}]-10=+6\,.
\label{eq:Rcharge2}
\ee
Therefore, from Eqs.~(\ref{eq:Rcharge1}) and (\ref{eq:Rcharge2}), one obtains 
\be
3R[\Phi]+2R[\Phi']=-6\,,
\label{eq:Rcharge3}
\ee
which is nothing but $R[\mathcal{O}_{\cancel{PQ}}^{(n=1)}]$.

This logic above tells us that the $R$-charge assignment inferred from the gauge-invariant $W_{\rm Yuk}$ for the hidden sector and the mixed anomaly free conditions automatically determine the $R$-charge of the operator in Eq.~(\ref{eq:O1}), irrespective of the detailed $R$-charge assignment in the hidden sector. With $R[\mathcal{O}_{\cancel{PQ}}^{(n=1)}]=-6$ as the model's prediction, there can be two contributions to the total $V$ in Eq.~(\ref{eq:VSUGRA}) to check for $\Delta\bar{\theta}_{\rm min}$. These are contributions coming from $(\partial W/\partial \Theta_{a})\times(W^{\dagger}/M_{P}^{2})(\partial K/\partial\Theta^{\dagger}_{a})$ ({\bf contribution 1}) and $(W/M_{P}^{2})(\partial K/\partial\Theta_{a})\times(W^{\dagger}/M_{P}^{2})(\partial K/\partial\Theta^{\dagger}_{a})$ ({\bf contribution 2}) where $\Theta_{a}=\Phi,\bar{\Phi},\Phi',\bar{\Phi}'$. Let us begin with the first case.\\

\underline{{\bf Contribution 1}}: If $N>8$ is chosen for $Z_{NR}$, $\mathcal{O}_{\cancel{PQ}}^{(n=1)}$ should be unavoidably multiplied by $(m_{3/2}/M_{P})^{4}$ to appear in the superpotential. Finally $\mathcal{O}_{\cancel{PQ}}^{(n=1)}$'s contribution to the scalar potential of the model in SUGRA reads
\be
V(\Phi,\Phi')\supset c_{1}m_{3/2}^{\dagger}M_{P}^{3}\left(\frac{m_{3/2}}{M_{P}}\right)^{4}\frac{\Phi^{3}\Phi'^{2}}{M_{P}^{5}}+
{\rm h.c.}\,.
\label{eq:VPhiPhi'}
\ee
This additional contribution to the axion potential causes the shift in the global minimum of the axion potential by
\be
\Delta\bar{\theta}_{\rm min}=3\times10^{-18}\times c_{1}\times\left(\frac{m_{3/2}}{10^{6}{\rm GeV}}\right)^{5}\times\left(\frac{v}{10^{12}{\rm GeV}}\right)^{5}\,.
\label{eq:thetamin1}
\ee
Hence, as long as we impose $Z_{NR}$ with $N>8$, $\mathcal{O}_{\cancel{PQ}}^{(n=1)}$ never spoils the Peccei-Quinn mechanism to solve the strong CP problem. 

We regard Eq.~(\ref{eq:thetamin1}) as a remarkable consequence of the model as TeV scale $m_{3/2}$ can be consistent with $\bar{\theta}_{\rm min}<10^{-10}$, which was not the case in \cite{Choi:2020vgb}. If one assumes a charge assignment resulting in $q+p\gtrsim12$~\cite{Fukuda:2018oco}, of course TeV scale $m_{3/2}$ can be allowed for $v=\mathcal{O}(10^{12}){\rm GeV}$. But such a large charge separation among $\Phi$ and $\Phi'$ is a bit artificial, lacking any field theoretic logical justification.

Then, does every $Z_{NR}$ with $N>8$ fully eliminate all the potential source of $\Delta\bar{\theta}_{\rm min}$ in the model? In answering this question, care must be taken not to miss\footnote{Considering the suppression due to $(m_{3/2}\mu/M_{P}^{2})$ is redundant because of $R[m_{3/2}\mu]=0$.} 
\begin{itemize}
    \item operators $(m_{3/2}/M_{P})^{\Delta}\mathcal{O}^{(n>1)}$ with $\Delta\in\mathbb{N}$
    \item the case where one of operators in Eq.~(\ref{eq:WPQviolating}) with $n>1$ appears in the superpotential without any suppression by powers of ($m_{3/2}/M_{P}$) by respecting $Z_{NR}$ on its own (type A1)
    \item the most dangerous operator multiplied by a powers of $(\mu/M_{P})$, e.g., $(\mu/M_{P})^{\alpha}\mathcal{O}_{\cancel{PQ}}^{(n=1)}$ with $\alpha\in\mathbb{N}$, which satisfies $R[(\mu/M_{P})^{\alpha}\mathcal{O}_{\cancel{PQ}}^{(n=1)}]=2\,\,{\rm mod}\,\,N$ (type A2)
\end{itemize}

For the first one, we find that $(\Delta,n)=(5,2),(4,3),(3,4),(2,5),(1,6)$ are safe enough combinations. This means that for each $n$, $\Delta$'s greater than the indicated ones are good enough for axion quality. Given this, we see that the choice satisfying $N\geq36$ makes us free of any dangerous $\Delta\bar{\theta}_{\rm min}$ from the operators $(m_{3/2}/M_{P})^{\Delta}\mathcal{O}^{(n>1)}$ with $\Delta\in\mathbb{N}$.

Concerning the second case, for instance, if $Z_{14R}$ is imposed, the operator in Eq.~(\ref{eq:WPQviolating}) with $n=2$ can contribute to the scalar potential as follows
\be
V(\Phi,\Phi')\supset c_{2}m_{3/2}^{\dagger}M_{P}^{3}\left(\frac{\Phi^{3}\Phi'^{2}}{M_{P}^{5}}\right)^{2}+
{\rm h.c.}\,.
\label{eq:VPhiPhi'2}
\ee
And this leads to 
\be
\Delta\bar{\theta}_{\rm min}=1.4\times c_{2}\times\left(\frac{m_{3/2}}{10^{6}{\rm GeV}}\right)\times\left(\frac{v}{10^{12}{\rm GeV}}\right)^{10}\,,
\label{eq:thetamin2}
\ee
which shows $m_{3/2}\lesssim10{\rm keV}$ is required for $v=10^{12}{\rm GeV}$. We notice that the gravitino cosmology can drastically change depending on $m_{3/2}$ and the choice of $m_{3/2}$ should be consistent with the null discovery of sparticles in the collider searches. Thus taking an arbitrary small enough $m_{3/2}$ is not allowed and rendering $\Delta\bar{\theta}_{\rm min}$ caused by the next leading contribution like Eq.~(\ref{eq:VPhiPhi'2}) small enough is non-trivial. 

We find that the unsuppressed $\mathcal{O}_{\cancel{PQ}}^{(n=7)}$ itself leads to $\Delta\bar{\theta}_{\rm min}=4\times10^{-20}$ whereas $\mathcal{O}_{\cancel{PQ}}^{(n<7)}$ does $\Delta\bar{\theta}_{\rm min}>10^{-10}$ for $m_{3/2}\leq10^{6}{\rm GeV}$ and $v\leq10^{16}{\rm GeV}$. Thus type A1 operators are not problematic for $Z_{NR}$ with $N\geq39$.

Lastly, apart from $m_{3/2}$, one should not forget the presence of the other spurion of $R$-symmetry in the model, i.e. the higgsino mass parameter $\mu$ with $R[\mu]=-2\,\,{\rm mod}\,\,N$ (see Appendix.~\ref{sec:appendixB}). For a $N>8$ for $Z_{NR}$, even if $\mathcal{O}_{\cancel{PQ}}^{(n=1)}$ does not appear in the superpotential, $(\mu/M_{P})\mathcal{O}_{\cancel{PQ}}^{(n=1)}$ can do if $N=10$. This operator contributes to the scalar potential as follows
\be
V(\Phi,\Phi')\supset c_{\mu}m_{3/2}^{\dagger}\mu M_{P}^{2}\left(\frac{\Phi^{3}\Phi'^{2}}{M_{P}^{5}}\right)+
{\rm h.c.}\,.
\label{eq:VPhiPhi'3}
\ee
And this leads to 
\be
\Delta\bar{\theta}_{\rm min}=4\times10^{16}\times c_{\mu}\times\left(\frac{m_{3/2}}{10^{6}{\rm GeV}}\right)\times\left(\frac{\mu}{10^{3}{\rm GeV}}\right)\times\left(\frac{v}{10^{12}{\rm GeV}}\right)^{5}\,.
\label{eq:thetamin3}
\ee
We find that $(\mu/M_{P})^{2}\mathcal{O}_{\cancel{PQ}}^{(n=1)}$ is also dangerous because, if allowed in the superpotential, it contributes to the scalar potential by
\be
V(\Phi,\Phi')\supset c_{\mu,2}m_{3/2}^{\dagger}\mu^{2} M_{P}\left(\frac{\Phi^{3}\Phi'^{2}}{M_{P}^{5}}\right)+
{\rm h.c.}\,.
\label{eq:VPhiPhi'4}
\ee
And it causes 
\be
\Delta\bar{\theta}_{\rm min}=18\times c_{\mu,2}\times\left(\frac{m_{3/2}}{10^{6}{\rm GeV}}\right)\times\left(\frac{\mu}{10^{3}{\rm GeV}}\right)^{2}\times\left(\frac{v}{10^{12}{\rm GeV}}\right)^{5}\,,
\label{eq:thetamin4}
\ee
Thus $Z_{10}$ and $Z_{12}$ must be also avoided for the choice of $Z_{NR}$. We find that $(\mu/M_{P})\mathcal{O}_{\cancel{PQ}}^{(n=5)}$ results in $\Delta\bar{\theta}_{\rm min}=10^{-11}$ whereas $(\mu/M_{P})\mathcal{O}_{\cancel{PQ}}^{(n<5)}$ does $\Delta\bar{\theta}_{\rm min}>10^{-10}$ for $\mu\leq10^{3}{\rm GeV}$, $m_{3/2}\leq10^{6}{\rm GeV}$ and $v\leq10^{16}{\rm GeV}$. This means that we need not worry about type A2 operators for $Z_{NR}$ with $N\geq34$.\\

\underline{{\bf Contribution 2}}: For a given $Z_{NR}$ with $N>8$, the most dangerous contributions from $(W/M_{P}^{2})(\partial K/\partial \Theta_{a})\times(W^{\dagger}/M_{P}^{2})(\partial K/\partial \Theta^{\dagger}_{a})$ include 
\be
V\supset\begin{cases}
    m_{3/2}^{\dagger}m_{3/2}\left(\frac{m_{3/2}}{M_{P}}\right)^{\beta}\frac{\mathcal{O}^{(n\geq1)}_{\cancel{PQ}}}{M_{P}}\,,\\
    m_{3/2}^{\dagger}m_{3/2}\frac{\mathcal{O}^{(n>1)}_{\cancel{PQ}}}{M_{P}}, & {\rm with}\,\, R[\mathcal{O}^{(n>1)}_{\cancel{PQ}}]=0\,\,{\rm mod}\,\,N\,\, ({\rm type\,\, B1})\,,\\
    m_{3/2}^{\dagger}m_{3/2}\left(\frac{\mu}{M_{P}}\right)^{\gamma}\frac{\mathcal{O}^{(n>1)}_{\cancel{PQ}}}{M_{P}}, &{\rm with}\,\,R[\mu^{p}\mathcal{O}^{(n>1)}_{\cancel{PQ}}]=0\,\,{\rm mod}\,\,N\,\, ({\rm type\,\, B2})\,,
  \end{cases}
\label{eq:Kahler2}
\ee
where $\beta,\gamma\in\mathbb{N}$.

For the first case in Eq.~(\ref{eq:Kahler2}), we should be aware that $(m_{3/2}/M_{P})^{\beta}\mathcal{O}_{\cancel{PQ}}^{(n\geq1)}/M_{P}$ ($\beta\in\mathbb{N}$) can appear in the Kahler potential together with its hermitian conjugate if $R[m_{3/2}^{\beta}\mathcal{O}_{\cancel{PQ}}^{(n\geq1)}]=0\,\,{\rm mod}\,\,N$. For example, applying to the most dangerous one $\mathcal{O}_{\cancel{PQ}}^{(n=1)}$, we can have
\be
K\supset \frac{c_{K1}}{M_{P}^{6}}(m_{3/2}^{3}\Phi^{3}\Phi'^{2}+{\rm h.c.})\,\,+\,\,\frac{\bar{c}_{K1}}{M_{P}^{6}}(m_{3/2}^{3}\Phi^{3}\bar{\Phi}'^{\dagger\,\,2}+{\rm h.c.})\,,
\label{eq:Kahler1}
\ee
where $c_{K1}$ and $\bar{c}_{K1}$ are dimensionless coefficients. Taking into account the canonically normalized kinetic terms in $K$, we expect Eq.~(\ref{eq:Kahler1}) results in $\Delta\bar{\theta}_{\rm min}$ equivalent to Eq.~(\ref{eq:thetamin1}) up to the dimensionless coefficients. Thus as far as there is sufficient suppression of Eq.~(\ref{eq:VPhiPhi'}), Eq.~(\ref{eq:Kahler1}) does not spoil the Peccei-Quinn mechanism. 

Next, paired with the hermitian conjugate, in Kahler potential there can be the operator $\mathcal{O}^{(n>1)}_{\cancel{PQ}}$ whose $R$-charge is 0 mod $N$ on its own (type B1 in Eq.~(\ref{eq:Kahler2})) or $R[\mu^{\beta}\mathcal{O}^{(n>1)}_{\cancel{PQ}}]=0\,\,{\rm mod}\,\,N$ (type B2 in Eq.~(\ref{eq:Kahler2})). Note that type B1 (B2) is analogous to type A1 (A2).

For operators belonging to type B1 (B2) to appear in the Kahler potential, $Z_{(N-2)R}$ is required when $Z_{NR}$ is needed for operators of type A1 (A2) to appear in the superpotential.\footnote{Put in another way, for example, if $Z_{14R}$, $Z_{16R}$, $Z_{18R}$ are required for appearance of $\mathcal{O}^{(n=2)}_{\cancel{PQ}}$, $(\mu/M_{P})\mathcal{O}^{(n=2)}_{\cancel{PQ}}$ and $(\mu/M_{P})^{2}\mathcal{O}^{(n=2)}_{\cancel{PQ}}$ in $W$, then $Z_{12R}$, $Z_{14R}$, $Z_{16R}$ are required for the same operators to appear in $K$.} Therefore, insofar as we choose $Z_{NR}$ that ensures sufficient suppression for operators of type A1 (A2), we need not worry about operators of the type B1 (B2) because of $N-2<N$.

In sum, it suffices to consider a choice of $Z_{NR}$ that makes {\bf contribution 1} suppressed enough. With that being said, a choice with $N\geq39$ would provide us with good enough axion quality as operators belonging to both type A1 and A2 (and thus type B1 and B2) get sufficiently suppressed. One may ask if some odd $N$s residing in $8<N\leq39$ can be working examples. $N=21$ can be an example. Now that we are armed with logically-supported freedom to choose whatever a large $N$ for $Z_{NR}$ we desire, we do not perform a further analysis to answer this question.

Though requiring $N$ to be greater than a certain threshold seems a bit strong condition, this is logically well-justified in our framework since different choice of $N$ for $Z_{NR}$ does not change any thing in the model (charge assignment and superpotential), never affects if the mixed anomaly free conditions for gauge symmetries are fulfilled, and does not require a large number of new colored fermions nor an exotic large $R$-charge. All these merits of our solution are attributable to the zero mixed anomaly condition
\be
\mathcal{A}_{NR2}=\mathcal{A}_{NR3}=0\,.
\ee
If one accepts a non-zero integer multiple of 
$N$ as a value of $\mathcal{A}_{NR2}=\mathcal{A}_{NR3}$, practically searching for viable scenarios featured by distinct $R$-charge assignments and new extra multiplets becomes very complicated~\cite{Harigaya:2013vja}. We are taking exactly the opposite point of view. This is the key point of this work. Thereby, our framework offers the high quality axion logically, fully guaranteeing consistency with any SUSY-breaking scale and the related SUSY phenomenologies.\footnote{It turns out that still this solution does not address the axion quality problem fully for $F_{a}\gtrsim\mathcal{O}(10^{16}){\rm GeV}$. We will get back to this issue in Sec.~\ref{sec:tension}.}

We conclude this section by commenting on a way to have stronger suppression than Eq.~(\ref{eq:thetamin1}). Recall that the power of gravitino mass in Eq.~(\ref{eq:VPhiPhi'}) is determined by Eq.~(\ref{eq:Rcharge3}). This means that one can have higher powers of $m_{3/2}$ than in Eq.~(\ref{eq:thetamin1}) if $\mathcal{A}_{NR2}=\mathcal{A}_{NR3}<-6$ can be realized. Perhaps in the models of quintessence axion~\cite{Fukugita:1994xx,Frieman:1995pm,Choi:1999xn} or fuzzy dark matter~\cite{Hui:2016ltb}, one needs higher suppression by powers of $(m_{3/2}/M_{P})$ than Eq.~(\ref{eq:VPhiPhi'}). In that case, for instance, one may add $k$-pairs of ${\bf 5}+{\bf 5}^{*}$ with $R[{\bf 5}{\bf 5}^{*}]=0$ resulting in $\mathcal{A}_{NR2}=\mathcal{A}_{NR3}=-6-2k$.

\subsection{Large Axion Potential from Small Size Instanton?}
\label{sec:model2}
In Sec.~\ref{sec:model2}, we introduced five sets of (${\bf 5},{\bf 5}^{*}$) as the minimum number of necessary fields for the vanishing mixed anomalies of $U(1)_{\rm B-L}$. One may wonder if this price we had to pay for $U(1)_{\rm B-L}$ gauge symmetry-assisted high quality of axion introduces another axion quality problem by triggering a significant small size instanton-induced contribution to the normal QCD axion potential.

Before integrating out the five sets of (${\bf 5},{\bf 5}^{*}$), i.e. before $U(1)_{\rm B-L}$ and $U(1)_{\rm PQ}$ breaking, the first beta function coefficient of 
$SU(3)_{c}$ becomes modified as
\be
b_{3,{\rm new}}=b_{3,{\rm MSSM}}+\frac{2}{3}T(\ytableausetup{textmode, centertableaux, boxsize=0.6em}
\begin{ytableau}
 \\
\end{ytableau})(N_{{\bf 5}}+N_{{\bf 5}^{*}})+\frac{1}{3}T(\ytableausetup{textmode, centertableaux, boxsize=0.6em}
\begin{ytableau}
 \\
\end{ytableau})(N_{{\bf 5}}+N_{{\bf 5}^{*}})\,,
\label{eq:b3}
\ee
where $N_{{\bf 5}}$ $(N_{{\bf 5}^{*}})$ is the number of new fields in the (anti) fundamental representation of $SU(5)$. The first and second new contributions in Eq.~(\ref{eq:b3}) are attributed to the new fermions and sfermions respectively. Now that we have $(N_{{\bf 5}},N_{{\bf 5}^{*}})=(5,5)$ and $b_{3,{\rm MSSM}}=-3$, the modified one-loop beta function coefficient of QCD becomes $b_{3,{\rm new}}=b_{3,{\rm MSSM}}+5=+2$.

Therefore for the energy regime above a mass scale $M_{{\bf 5}}$ of $N_{{\bf 5}}$ and $N_{{\bf 5}^{*}}$, the supersymmetric QCD in our framework becomes a non-asymptotic free theory while it remains asymptotic free below $M_{{\bf 5}}$. The smaller $M_{{\bf 5}}$ is expected to give rise to a large $g_{s}$ at $M_{P}$, and thus will make the axion potential induced by the instanton of the size $\rho\sim M_{P}^{-1}$ larger. Does this variation of the axion potential due to the change of high energy behavior of QCD cause a dangerous $\Delta\bar{\theta}_{\rm min}$? The validity of our mechanism for addressing the axion quality problem will depend on an answer to this question. So, this section is dedicated to making the answer and proving the validity of the mechanism. 

Envisioning the high scale SUSY-breaking, we consider the gravity-mediated SUSY breaking scenarios.\footnote{Obviously the SUSY-breaking effect will appear in the visible sector after integrating out the SUSY-breaking mediation messengers. For other SUSY-breaking mediations like gauge meditation, usually the messenger mass scale satisfies $M_{\rm mess}<\!\!<M_{P}$. As the instantons with the size $\rho<M_{\rm mess}^{-1}$ do not contribute to the axion potential, for a given non-asymptotic $SU(3)_{c}$ gauge theory, $\Delta\bar{\theta}_{\rm min}$ caused by small size instantons is less dangerous in SUSY-breaking mediation scenarios other than the gravity mediation.} In supersymmetric theories, the leading contribution to the axion potential arises from the Kahler potential, which contains at least two powers of suppression factor $\rho m_{\rm soft}$~\cite{Choi:1998ep}. Here $\rho$ is an instanton size of interest and $m_{\rm soft}$ is a soft SUSY-breaking mass. For our case, $\rho$ is given by the inverse of the reduced Planck scale $M_{P}^{-1}$. 

The crude, but the most aggressive estimate of the magnitude of the axion potential induced by the instanton of the size $\rho\sim M_{P}^{-1}$ is given by
\be
\Delta V(\theta+\delta)\simeq e^{-\frac{2\pi}{\alpha_{s}(M_{P})}-i\bar{\theta}-i\delta}\times m_{3/2}^{2}M_{P}^{2}\times\left(\frac{F_{a}}{M_{P}}\right)^{T(\ytableausetup{textmode, centertableaux, boxsize=0.6em}
\begin{ytableau}
 \\
\end{ytableau})(N_{{\bf 5}}+N_{{\bf 5}^{*}})}+{\rm h.c.}\,,
\label{eq:instantonV}
\ee
where $g_{s}^{2}\equiv4\pi\alpha_{s}$ and $m_{\rm soft}=\mathcal{O}(m_{3/2})$ were used, and $F_{a}$ characterizes the mass scale for the newly added multiplets ${\bf 5}$ and ${\bf 5}^{*}$. Note that there exist $2T({\bf R})$ fermion zero modes in the representation ${\bf R}$ of $SU(3)_{c}$, which requires $T({\bf R})$ mass insertions for closing the fermion zero modes. This explains the last factor in Eq.~(\ref{eq:instantonV}) which reflects closing the fermion zero modes of new multiplets. 

We assume the phase shift by $\delta=\mathcal{O}(1)$ that might arise from a CP-violating source in a UV physics. For example, a complex coefficient of the dimension six four fermion gauge invariant operators can be a source of $\delta\neq0$~\cite{Demirtas:2021gsq}. There can be several concrete UV physics constructions where CP is not conserved at high energy~\cite{Dine:1986bg}. Thus, it seems more reasonable to assume $\delta\neq0$.\footnote{Already within the SM, the non-vanishing shift $\Delta\bar{\theta}_{\rm min}=\mathcal{O}(10^{-19})$ is caused via the CP-violation in the electroweak sector~\cite{Georgi:1986kr}.}

We present the estimate in Eq.~(\ref{eq:instantonV}) as the largest possible magnitude of $\Delta V(\theta+\delta)$, and $\Delta V(\theta+\delta)$ in Eq.~(\ref{eq:instantonV}) may not be invariant under $Z_{NR}$. For the proper $Z_{NR}$ invariance, $\Delta V(\theta+\delta)$ should be multiplied by powers of $(m_{R}/M_{P})$ with $m_{R}$ collectively denoting VEVs of fields carrying a non-zero $R$-charge. Furthermore, dimensionless coupling constants used for closing fermion zero modes may also appear in $\Delta V(\theta+\delta)$, giving a further suppression. Then the actual $Z_{NR}$ invariant $\Delta V(\theta+\delta)$ is expected to be smaller than Eq.~(\ref{eq:instantonV}) in magnitude. Thus if we can show that $\Delta\bar{\theta}_{\rm min}$ due to Eq.~(\ref{eq:instantonV}) is sufficiently suppressed to satisfy $\Delta\bar{\theta}_{\rm min}<10^{-10}$, our framework for addressing the axion quality problem remains intact. 

Now we are in position to show that Eq.~(\ref{eq:instantonV}) is not at all dangerous to cause a significant $\Delta\bar{\theta}_{\rm min}$ as large as $10^{-10}$. To this end, we rewrite gauge coupling appearing in the exponent of the instanton amplitude as
\ba
\frac{2\pi}{\alpha_{s}(M_{P})}&=&\frac{2\pi}{\alpha_{s}(F_{a})}-b_{3,{\rm new}}\log\left(\frac{M_{P}}{F_{a}}\right)\cr\cr
\frac{2\pi}{\alpha_{s,{\rm MSSM}}(F_{a})}&=&\frac{2\pi}{\alpha_{s,{\rm MSSM}}(M_{Z})}-b_{3,{\rm MSSM}}\log\left(\frac{F_{a}}{M_{Z}}\right)
\,,
\label{eq:alphac1}
\ea
where $4\pi\alpha_{s}\equiv g_{s}^{2}$ with $g_{s}$ $SU(3)_{c}$ gauge coupling with the new field contents in Table.~\ref{table:qn2}, $4\pi\alpha_{s,{\rm MSSM}}\equiv g_{s,{\rm MSSM}}^{2}$ with $g_{s,{\rm MSSM}}$ $SU(3)_{c}$ gauge coupling within the MSSM and $M_{Z}\simeq91.2{\rm GeV}$ is the $Z$-boson mass. Note that $\alpha_{s}(F_{a})=\alpha_{s,{\rm MSSM}}(F_{a})$ holds true as the boundary condition. Substituting the second equation into the first equation in Eq.~(\ref{eq:alphac1}), we obtain
\be
\frac{2\pi}{\alpha_{s}(M_{P})}=\frac{2\pi}{\alpha_{s,{\rm MSSM}}(M_{Z})}-b_{3,{\rm MSSM}}\log\left(\frac{M_{P}}{M_{Z}}\right)-T(\ytableausetup{textmode, centertableaux, boxsize=0.6em}
\begin{ytableau}
 \\
\end{ytableau})(N_{{\bf 5}}+N_{{\bf 5}^{*}})\log\left(\frac{M_{P}}{F_{a}}\right)\,.
\label{eq:alphac2}
\ee
Putting Eq.~(\ref{eq:alphac2}) in Eq.~(\ref{eq:instantonV}) gives us 
\be
\Delta V(\theta)\simeq e^{-\frac{2\pi}{\alpha_{s,{\rm MSSM}}(M_{P})}-i\bar{\theta}-i\delta}\times m_{3/2}^{2}M_{P}^{2}+{\rm h.c.}\,.
\label{eq:instantonV2}
\ee

Remarkably, this result tells us that the axion potential contribution from the small size instanton of the modified QCD is equivalent to that induced by the same size instanton of the QCD in the MSSM at the one-loop level. Put in another way, the axion potential induced by the small size instanton is insensitive to UV modification to RGE of QCD gauge coupling we have in the model.\footnote{The insensitivity of the axion potential to the modification of the RGE of a non-Abelian gauge theory was observed in \cite{Nomura:2000yk} and used in \cite{Ibe:2018ffn,Choi:2019jck,Choi:2021aze} in the context of electroweak (EW) quintessence axion.} Therefore, Eq.~(\ref{eq:instantonV2}) removes the concern about too a large $\Delta\bar{\theta}_{\rm min}$ due to the modified QCD gauge coupling RGE. 

By solving RGE for $\alpha_{s,{\rm MSSM}}$, we can obtain its value evaluated at $M_{P}$. Evaluation of $\Delta\bar{\theta}_{\rm min}$ due to Eq.~(\ref{eq:instantonV2}) is estimated to be
\be
\Delta\bar{\theta}_{\rm min}\simeq2\times10^{-12}\times\left(\frac{m_{3/2}}{10^{10}{\rm GeV}}\right)^{2}\,.
\label{eq:thetamin3}
\ee

Therefore, from Eq.~(\ref{eq:thetamin3}), we conclude that the modification to QCD with the new colored fields in Table.~\ref{table:qn2} never causes a dangerous $\Delta\bar{\theta}_{\rm min}$ as long as $m_{3/2}\lesssim10^{10}{\rm GeV}$.

\section{Tension with QCD String Axion}
\label{sec:tension}
\setcounter{equation}{0}
In Sec.~\ref{sec:app}, we showed how the gauged $U(1)_{\rm B-L}$ and the discrete $R$-symmetry $Z_{NR}$ can protect the Peccei-Quinn mechanism from being spoiled by contributions to $\Delta\bar{\theta}_{\rm min}>\!\!>1$ from higher dimensional operators. As the main result of the framework, remarkable is that Eq.~(\ref{eq:thetamin1}) shows that essentially the gravitino mass as large as $m_{3/2}=\mathcal{O}(10^{4}){\rm GeV}$ can be consistent with $\Delta\bar{\theta}_{\rm min}\leq10^{-10}$ for $F_{a}=\mathcal{O}(10^{15}){\rm GeV}$. 

This result is very appealing, given the null observation of any sparticles in the LHC. Moreover, if the high quality axion requires a somewhat lower regime of $m_{3/2}$ than TeV scale, there can be a cosmological danger to spoil the Big Bang Nucleosythesis (BBN) due to the presence of particles with mass $\mathcal{O}(m_{3/2})$. Particularly those particles which have GeV scale or lower mass, but very weakly interact with other particle contents in the model can be potentially dangerous by undergoing the decay into radiation after BBN era. This will destroy the existing primordial light elements, causing inconsistency with the experimentally measured primordial light element abundance.

This concern actually applies to the unavoidable particle contents of the model, saxion ($S$)~\cite{Fox:2004kb}. Saxion is the real part of the scalar component of axion supermultiplet in supersymmetric theory and it obtains the mass of $\mathcal{O}(m_{3/2})$ in SUGRA. The strength of its interaction with other particles are characterized by $F_{a}^{-1}$ just like axion. Because the high axion quality does not allow for simultaneous largeness for $m_{3/2}$ and $F_{a}$ as can seen in Eq.~(\ref{eq:thetamin1}), a problematic low $m_{3/2}$ for $F_{a}$ as large as $\mathcal{O}(10^{16}){\rm GeV}$ may be required for the high quality axion in the model. If the required $m_{3/2}$ is too low for $S$ to decay prior to BBN, then the model cannot fully resolve the axion quality for $F_{a}\gtrsim\mathcal{O}(10^{16}){\rm GeV}$. This section is dedicated to explore the range of $F_{a}$ wherein our resolution to the axion quality problem can be consistent with BBN.

Given the axion supermultiplet coupled to the gluon supermultiplet, the saxion decay to a pair of gluons becomes the main decay channel for the saxion. The decay rate of the process reads
\be
\Gamma_{\sigma\rightarrow gg}=\frac{\alpha_{s}^{2}}{64\pi^{3}}\frac{m_{S}^{3}}{F_{a}^{2}}\,.
\label{eq:decayrate}
\ee
We can infer the temperature of the thermal bath when the decay happens by comparing $\Gamma_{\sigma\rightarrow gg}$ to the Hubble expansion rate during the radiation dominated era. In doing so, we obtain
\be
T_{S{\rm dec}}\simeq26{\rm MeV}\times\left(\frac{g_{*\rho}(a_{S{\rm dec}})}{10.75}\right)^{-1/4}\left(\frac{m_{S}}{100{\rm TeV}}\right)^{3/2}\left(\frac{F_{a}}{10^{16}{\rm GeV}}\right)^{-1}\,,
\label{eq:THR2}
\ee
where $m_{S}$ is the saxion mass and $g_{*\rho}(a_{S{\rm dec}})$ is the number of relativistic degrees of freedom at the time of $S$-decay.\footnote{Saxion can couple to quarks via, for example, $K\supset(A+A^{\dagger})Q^{\dagger}Q/F_{a}$. But the decay of saxion to a pair of quarks is subdominant in comparison to Eq.~(\ref{eq:decayrate}) as the corresponding decay rate is $\propto m_{S}(m_{Q}/F_{a})^{2}$.}

\begin{figure}[t]
\centering
\hspace*{-5mm}
\includegraphics[width=0.7\textwidth]{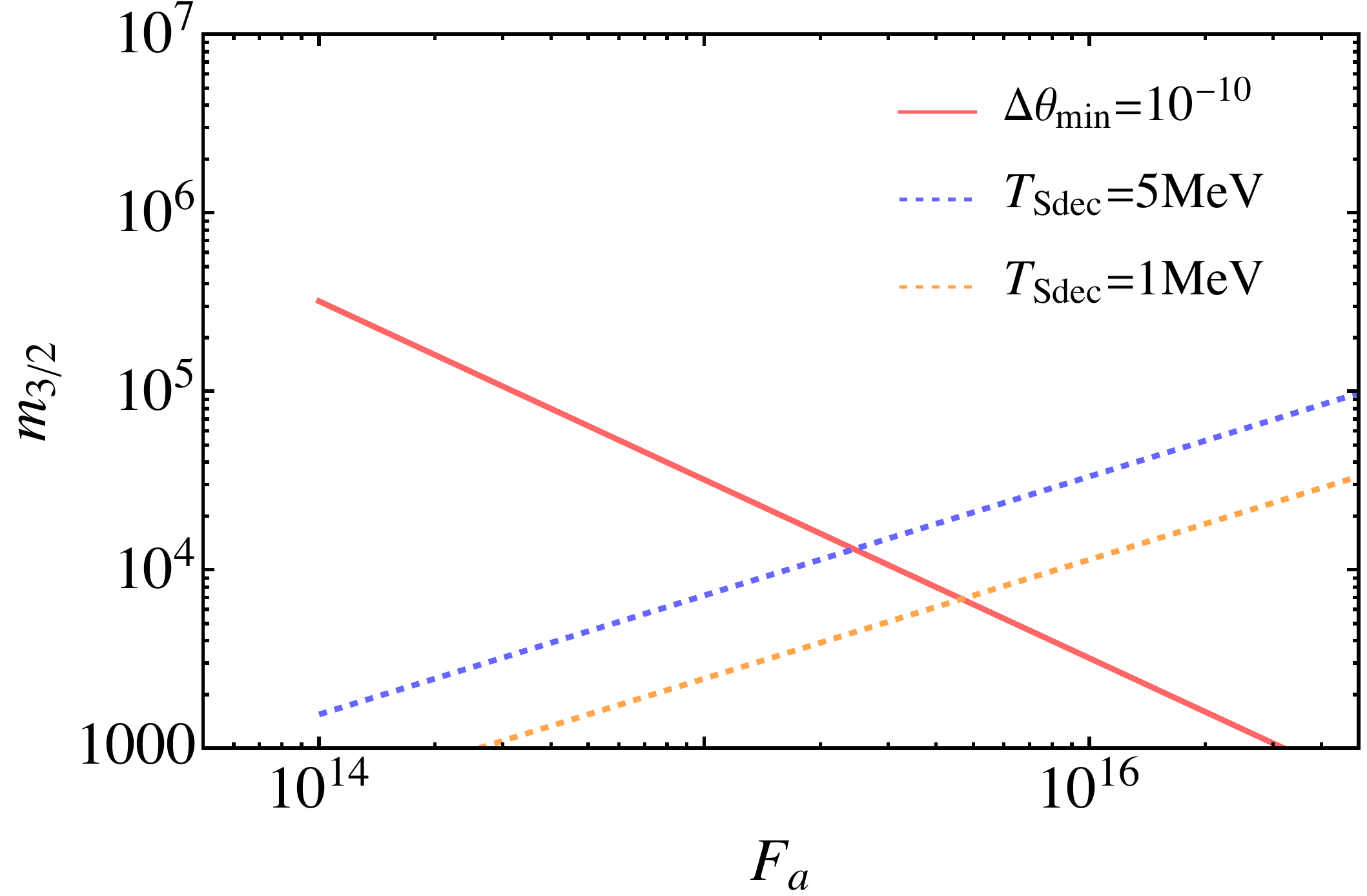}
\caption{Plot showing the collection of the point ($F_{a},m_{3/2}$) (1) satisfying $\Delta\bar{\theta}_{\rm min}=10^{-10}$ (red solid) based on Eq.~(\ref{eq:thetamin1}) and (2) producing $T_{S{\rm dec}}=5{\rm MeV}$ (blue dashed) and 1MeV (yellow dashed) based on Eq.~(\ref{eq:THR2}). For both cases, the parameter space below the lines makes $\Delta\bar{\theta}_{\rm min}$ and $T_{S{\rm dec}}$ smaller than the indicated values.}
\vspace*{-1.5mm}
\label{fig1}
\end{figure}

In Fig.~\ref{fig1}, we present the plot showing the collection of the points ($F_{a},m_{3/2}$) satisfying $\Delta\bar{\theta}_{\rm min}=10^{-10}$ (red solid), and $T_{S{\rm dec}}=5{\rm MeV}$ (blue dashed) and 1MeV (yellow dashed). For the parameter space below each line, $\Delta\bar{\theta}_{\rm min}$ and $T_{S{\rm dec}}$ are smaller than the indicated values. As a lower bound of the safe $T_{S{\rm dec}}$, we take $\mathcal{O}(1)$ MeV.

Indeed, reflecting our concern, for $F_{a}=\mathcal{O}(10^{16}){\rm GeV}$, if $m_{S}\simeq m_{3/2}$, we see that $m_{3/2}$ required for the high quality axion renders the saxion too light to complete the decay before BBN era. This means essentially our proposal for solving the axion quality problem cannot apply to $F_{a}$ regimes where the red line is below the dashed lines. There, the cosmological consistency requires $m_{3/2}$ above the dashed lines for a fixed $F_{a}$ and thus another way of solving axion quality problem is required.\footnote{In the case where $m_{S}\gtrsim10m_{3/2}$ holds true, our solution to the axion quality problem can still apply to the QCD string axion with $F_{a}\simeq10^{16}{\rm GeV}$, still being consistent with BBN constraint.}

Nevertheless, we emphasize that our model still makes $F_{a}$ as large as $\mathcal{O}(10^{15}){\rm GeV}$ consistent with the measurement of the primordial element abundance. In trials to address the axion quality problem in SUGRA, it is still non-trivial to achieve the cosmological consistency by allowing for TeV scale $m_{3/2}$ (for instance, see \cite{Fukuda:2018oco,Choi:2020vgb} where a problematic low enough $m_{3/2}$ is needed for $F_{a}$ beyond the QCD axion window).

\section{Conclusion and Outlook}
\label{sec:conclusion}
In this paper, we suggested a complete solution to the axion quality problem based on symmetries which have good motivations in BSM physics for other purposes than the strong CP problem. The symmetries we invoked are supersymmetry, the gauged $U(1)_{\rm B-L}$ and a gauged discrete $Z_{NR}$ symmetry with a sufficiently large $N$.

With $U(1)_{\rm PQ}$ symmetry for the axion solution to the strong CP problem interpreted as an accidental symmetry protected by the gauged $U(1)_{\rm B-L}$ to some extent, the axion quality problem reduces to sufficient suppression of the operators given in Eq.~(\ref{eq:WPQviolating}). This is the point where $R$-symmetry is involved in our solution.

Ideally, as we discussed in Sec.~\ref{sec:suppress}, a gauged $U(1)_{R}$ is the best option for addressing the axion quality problem. But as it is practically very difficult to have a gauged $U(1)_{R}$, we asked if an alternative gauged $Z_{NR}$ can effectively serve as $U(1)_{R}$ in the axion quality problem. This requires a large $N\geq\mathcal{O}(10)$ choice.

Provided the mixed anomalies of a discrete symmetry with the MSSM gauge groups $SU(2)_{L}$ and $SU(3)_{c}$ are 0 mod $N$ within the MSSM, we must introduce either of additional colored fermions as many as $N$ carrying $U(1)_{\rm PQ}$ or few fermions carrying abnormal large $R$-charges. As a matter of fact, this is the case for a usual discrete symmetry $Z_{N}$.

But for a discrete $R$-symmetry $Z_{NR}$, the mixed anomalies $\mathcal{A}_{NR2}$ and $\mathcal{A}_{NR3}$ are not $0$ mod $N$ within the MSSM. This fact can make $Z_{NR}$ distinguished clearly from a discrete non $R$-symmetry $Z_{N}$'s. As $\mathcal{A}_{NR2}=\mathcal{A}_{NR3}\neq0$ mod $N$, one can introduce few fermions contributing to $\mathcal{A}_{NR2}$ and $\mathcal{A}_{NR3}$ carrying reasonable $R$-charges to achieve $\mathcal{A}_{NR2}=\mathcal{A}_{NR3}=0$. Then whatever $N$ is considered, there is no need to introduce many new fields and large $R$-charges (see Sec.~\ref{sec:suppress}). 

In accordance with this mechanism, we considered the situation where $\mathcal{A}_{NR2}=\mathcal{A}_{NR3}=0$ is realized in the full theory. We emphasize that this is the most important aspect of this work. Thanks to this, our model could enjoy the logically well-justified a large $N$ choice for $Z_{NR}$ and become fully free of any source of $\Delta\bar{\theta}_{\rm min}$. Consequently, differing from other solutions to axion quality problem in supersymmetric models, Peccei-Quinn mechanism can successfully operate even without needing small enough $m_{3/2}$ and $F_{a}$. In our framework, there is always a unavoidable $\Delta\bar{\theta}_{\rm min}$ from Eq.~(\ref{eq:thetamin1}) and thus non-vanishing neutron electric dipole moment is expected. The constraint on $(m_{3/2},F_{a})$ from $\Delta\bar{\theta}_{\rm min}$ is shown in Fig.~\ref{fig1}.

We believe that our mechanism for having a logically well-justified $Z_{NR}$ with $N\geq\mathcal{O}(10)$ can be further applied to other problems in particle physics model building. Whenever there is a need to suppress high dimensional operators, one can impose the zero mixed anomaly conditions of $Z_{NR}$. This will help the model remain minimal, suppressing all the unwanted non-renormalizable operators.

\section*{Acknowledgments}
G.C. would like to thank Emilian Dudas for the useful discussion. G.C. would like to acknowledge the Mainz Institute for Theoretical Physics (MITP) of the Cluster of Excellence PRISMA+(Project ID 39083149), for its hospitality and its partial support during the completion of this work. T. T. Y. is supported in part by the China Grant for
Talent Scientific Start-Up Project and by Natural Science Foundation of China (NSFC) under grant No. 12175134 as well as by World Premier International Research Center Initiative (WPI Initiative), MEXT, Japan.

\appendix
\section{Mixed Anomaly for $Z_{N}$ within MSSM} 
\label{sec:appendixA}
Let us denote a $Z_{N}$ charge of a field $X$ by $Q_{N}[X]$. Suppose the Higgsino mass parameter satisfies $Q_{N}[\mu]=0$. If $Q_{N}[H_{u}]=x$, then $Q_{N}[H_{d}]=Nm_{1}-x$ should hold for an integer $m_{1}\in\mathbb{Z}$. On the other hand, the first two Yukawa couplings in Eq.~(\ref{eq:MSSMW}) give the conditions $Q_{N}[{\bf 10}{\bf 10}H_{u}]=Nm_{2}$ and $Q_{N}[{\bf 10}\,{\bf 5}^{*}H_{d}]=Nm_{3}$ with $m_{2},m_{3}\in\mathbb{Z}$. 

Combining these conditions, one finds
\be
Q_{N}[{\bf 10}]=\frac{N}{2}m_{2}-\frac{x}{2}\quad,\quad Q_{N}[{\bf 5}^{*}]=\frac{N}{2}(2m_{3}-m_{2}-2m_{1})+3\frac{x}{2}\,.
\ee

Given $Q_{N}[{\bf 10}]$, $Q_{N}[{\bf 5}^{*}]$, $Q_{N}[H_{u}]$, and $Q_{N}[H_{d}]$, we find the mixed anomalies of $Z_{N}-[SU(3)_{c}]^{2}$
\be
\underbrace{3}_{3\,\, {\rm generation}}\times(3Q_{N}[{\bf 10}]+Q_{N}[{\bf 5}]^{*})=3\times\frac{N}{2}(2m_{3}+2m_{2}-2m_{1})=0\,\,{\rm mod}\,\,N\,,
\ee
and of $Z_{N}-[SU(2)_{L}]^{2}$
\be
\underbrace{3}_{3\,\, {\rm generation}}\times(\underbrace{3}_{\rm color}Q_{N}[{\bf 10}]+Q_{N}[{\bf 5}]^{*})+Nm_{1}=3\times\frac{N}{2}(2m_{3}+2m_{2})-2Nm_{1}=0\,\,{\rm mod}\,\,N\,.
\ee

Thus both mixed anomalies are 0 mod $N$.

\section{Mixed Anomaly for $Z_{NR}$ within MSSM} 
\label{sec:appendixB}
If $R[H_{u}]=x$, then $R[H_{d}]=N\ell_{1}-x+2-R[\mu]$ should hold for an integer $\ell_{1}\in\mathbb{Z}$. On the other hand, the first two Yukawa couplings in Eq.~(\ref{eq:MSSMW}) give the conditions $R[{\bf 10}{\bf 10}H_{u}]=N\ell_{2}+2$ and $R[{\bf 10}\,{\bf 5}^{*}H_{d}]=N\ell_{3}+2$ with $\ell_{2},\ell_{3}\in\mathbb{Z}$. 

Combining these conditions, one finds
\be
R[{\bf 10}]=\frac{N}{2}\ell_{2}-\frac{x}{2}+1\quad,\quad R[{\bf 5}^{*}]=\frac{N}{2}(2\ell_{3}-\ell_{2}-2\ell_{1})+3\frac{x}{2}-1+R[\mu]\,.
\ee

Given $R[{\bf 10}]$, $R[{\bf 5}^{*}]$, $R[H_{u}]$, and $R[H_{d}]$, Eq.~(\ref{eq:mixedanomalyfree}) yields the mixed anomalies of $Z_{N}-[SU(3)_{c}]^{2}$
\be
\mathcal{A}_{NR3}=6+\underbrace{3}_{3\,\, {\rm generation}}\times(3R[{\bf 10}]+R[{\bf 5}]^{*}-4)=3N(\ell_{3}+\ell_{2}-\ell_{1})+3R[\mu]\,,
\label{eq:ANR3}
\ee
and of $Z_{N}-[SU(2)_{L}]^{2}$
\ba
\mathcal{A}_{NR2}&=&4+\underbrace{3}_{3\,\, {\rm generation}}\times(\underbrace{3}_{\rm color}R[{\bf 10}]+R[{\bf 5}]^{*}-4)+R[
H_{u}H_{d}]-2\cr\cr&=&-2+N(3\ell_{3}+3\ell_{2}-2\ell_{1})+2R[\mu]\,.
\label{eq:ANR2}
\ea

The difference between these two mixed anomalies reads
\be
\mathcal{A}_{NR3}-\mathcal{A}_{NR2}=2+R[\mu]-N\ell_{1}=4-R[H_{u}H_{d}]\,.
\label{eq:ANR32}
\ee

From Eqs.~(\ref{eq:ANR3}), (\ref{eq:ANR2}) and (\ref{eq:ANR32}), requiring $Z_{NR}$ to be free of the mixed anomalies tells us that $R[H_{u}H_{d}]=4\,\,{\rm mod}\,\,N$. In other words, $R[\mu]=N\ell_{4}-2$ with $\ell_{4}\in\mathbb{Z}$, which in turn results in
\be
\mathcal{A}_{NR3}=3N(\ell_{3}+\ell_{2}-\ell_{1}+\ell_{4})-6\quad,\quad\mathcal{A}_{NR2}=N(3\ell_{3}+3\ell_{2}-2\ell_{1}+2\ell_{4})-6\,.
\ee

In the case of $\ell_{1}=\ell_{4}$, $\mathcal{A}_{NR2}=\mathcal{A}_{NR3}$ holds true and both are -6 mod $N$. In the other case of $\ell_{1}\neq\ell_{4}$, both are still -6 mod $N$, but $\mathcal{A}_{NR2}\neq\mathcal{A}_{NR3}$.

\bibliography{main}
\bibliographystyle{jhep}

\end{document}